\UseRawInputEncoding

\documentclass {article} 

\usepackage{citesort}

\begin{document}   

\author{J. Brian Pitts \\ University of Cambridge,  University of Lincoln, and University of South Carolina     \\ jbp25@cam.ac.uk, bpitts@lincoln.ac.uk, JP84@mailbox.sc.edu  }

\date{ }

\sloppy

\title{Change in Hamiltonian General Relativity with Spinors }

\maketitle


\abstract{ In General Relativity in Hamiltonian form, change has seemed to be missing, defined only asymptotically, or otherwise obscured at best, because the Hamiltonian is a sum of first-class constraints and a boundary term and thus supposedly generates gauge transformations.  By construing change as essential time dependence (lack of a time-like Killing vector field), one can find change locally in vacuum GR in the Hamiltonian formulation just where it should be. But what if spinors are present?

This paper is motivated by the tendency in space-time philosophy tends to slight  fermionic/spinorial matter, the tendency in  Hamiltonian GR to  misplace changes of time coordinate, and the tendency in treatments of the Einstein-Dirac equation to include a gratuitous local Lorentz gauge symmetry along with the physically significant coordinate freedom.  Spatial dependence is dropped in most of the paper, both restricting the physical situation and largely fixing the spatial coordinates.  In the interest of including all and only the coordinate freedom, the Einstein-Dirac equation is investigated using the Schwinger time gauge and Kibble-Deser symmetric triad condition are employed as  a $3+1$ version of the DeWitt-Ogievetsky-Polubarinov nonlinear group realization formalism that dispenses with a tetrad and local Lorentz gauge freedom. Change is the lack of a time-like stronger-than-Killing field for which the Lie derivative of the metric-spinor complex vanishes.  An appropriate $3+1$-friendly form of the Rosenfeld-Anderson-Bergmann-Castellani gauge generator $G$, a tuned sum of first class-constraints, is shown to change the canonical Lagrangian by a total derivative, implying the preservation of Hamilton's equations.  Given the essential presence of second-class constraints with spinors and their lack of resemblance to a gauge theory (unlike, say, massive photons), it is useful to have an explicit physically interesting example.  
This gauge generator implements changes of time coordinate for solutions of the equations of motion, showing that the gauge generator makes sense even with spinors.   }



\tableofcontents

\section{Introduction}

\subsection{Hamiltonian Change Seems Missing but Lagrangian Change Is Not}  

It has been argued  that General Relativity, at least in Hamiltonian form, lacks change, has change only asymptotically and hence only for certain topologies, or appears to lack change with no clear answer in sight (\emph{e.g.}, \cite{AndersonChange,IshamTime,BelotEarman,EarmanMcTaggart,RicklesTimeStructureQG,HuggettWuthrichTimeQG}).   Such a conclusion calls to mind earlier philosophical puzzles, whether ancient (the paradoxes of Zeno, whom James Anderson mentions \cite{AndersonRoyaumont,AndersonChange}, and the views of Parmenides, whom Kucha\v{r} mentions \cite{KucharCanonical93}) or modern (the argument conclusion that real time requires something contradictory and hence is impossible by McTaggart \cite{McTaggart}, mentioned in a memorable philosophical exchange \cite{EarmanMcTaggart,MaudlinMcTaggart}). 
Efforts to find deep philosophical lessons in the constrained Hamiltonian formalism \cite{EarmanMcTaggart,EarmanOde} have been  resisted 
 \cite{MaudlinMcTaggart,HealeyGRchangelessincoherent}, but without adequate diagnosis until recently \cite{PonsDirac,BarbourFosterPrimary,FirstClassNotGaugeEM,GRChangeNoKilling,ObservablesEquivalentCQG}. 
However, if one defines change  as essential time dependence, which comes to the lack of a time-like Killing vector field in vacuum GR  \cite[p. 355]{Ohanian}, then Hamiltonian GR has change exactly where non-Hamiltonian GR has change (restricting attention to space-times that admit a Hamiltonian treatment) \cite{GRChangeNoKilling}.  These results build upon the \emph{c.} 1980s+ Lagrangian-equivalent reforming literature  that has sought to recover Hamiltonian-Lagrangian equivalence and the freedom to change time coordinates  \cite{MukundaSymmetries,MukundaGaugeGenerator,SalisburySundermeyerEinstein,PonsSalisburyShepley,ShepleyPonsSalisburyTurkish,PonsSalisbury,PonsReduce,SuganoExtended,PonsDirac,SuganoGeneratorQM,SuganoGaugeVelocityI,SuganoGaugeGenerator}.  The story of how such equivalence and temporal coordinate freedom were lost in the 1950s-60s is starting to be written \cite{SalisburySyracuse1949to1962,BergmannObservables}.

Just as it was convenient to discard spatial dependence in order to simplify both the technical and conceptual aspects of the problem to uncover change and temporal coordinate freedom in vacuum GR \cite{GRChangeNoKilling}, it is useful to impose a similar simplification in treating the Einstein-Dirac equation, GR coupled to spinorial matter.  
This problem is suggested as a step toward addressing several unhelpful tendencies in various literatures.  First, conceptual reflection about space-time and GR (whether by physicists or by philosophers) often pays little attention to spinor fields, apart from global technical questions in certain quarters.  While admittedly one does not encounter manifestly spinorial-behaving matter very readily, the fact that the great bulk of matter (electrons, quarks, neutrinos\ldots) is represented in quantum field theory by (almost-anticommuting) spinor fields should suffice to make spinorial matter a prominent part of foundational reflection on space-time.  This paper can be viewed as a step to that end, though the spinors are assumed here to commute if it matters.

Second, during the later 1950s-60s it became common in Hamiltonian GR to discard a spatio-temporal viewpoint that retained the freedom to change time coordinates \cite{DiracHamGR,SalisburySyracuse1949to1962,BergmannObservables}, thus obscuring one of the most conceptually interesting features of GR; this bargain was viewed as an aid to quantization. 
 This move contrasts with the spatio-temporal coordinate freedom in the earlier Hamiltonian work by Rosenfeld and by Anderson and Bergmann \cite{RosenfeldQG,AndersonBergmann}. Mathematical equivalence between the Hamiltonian and Lagrangian formalisms was  discarded in favor of a supposed physical equivalence for ``observables''---but the definition of observables was not successfully mathematically grounded.  Since \emph{c.} 1980 a reforming literature has aimed to recover spatio-temporal coordinate freedom in GR and mathematical equivalence between the Hamiltonian and Lagrangian formulations (\emph{e.g.}, \cite{MukundaSymmetries,MukundaGaugeGenerator,SalisburySundermeyerEinstein,PonsSalisburyShepley,ShepleyPonsSalisburyTurkish,PonsSalisbury,PonsReduce,SuganoExtended,PonsDirac,SuganoGeneratorQM,SuganoGaugeVelocityI,SuganoGaugeGenerator}).  Recently it was shown how to bootstrap a definition of observables using equivalent formulations of massive theories, one without gauge freedom (so everything is observable) and one with gauge freedom \cite{ObservablesEquivalentCQG,ObservablesLSEFoP}.  It turns out that observables are basically tensor fields (or more generally geometric objects \cite{Nijenhuis,Schouten,Anderson} including connections or any set of components with a coordinate transformation rule) that are invariant under any internal gauge symmetries present \cite{ObservablesEinsteinMaxwellFoP}; thus observables are merely covariant, not invariant, under coordinate transformations.  The space-time metric and its concomitants therefore qualify as observables, which are local and change, contrary to the view that observables are global and constants of the motion.

A third unhelpful tendency is the belief that coupling spinors to gravity in accord with GR requires an orthonormal basis and hence local Lorentz ($O(3,1)$) freedom. 
Weyl and Cartan took themselves to  have proven this claim in 1929 \cite{WeylGravitationElectron,WeylElektronGravitation,WeylRice} or a bit later \cite[p. 151]{CartanSpinor}.  But if one compares Weyl's  theorem with his gloss of it, one sees that the gloss is considerably stronger; it does not follow from the theorem without additional assumptions, assumptions which were identified by Ogievetsky and Polubarinov in the 1960s \cite{OPspinor,OP,PittsSpinor}.  Weyl and Cartan assumed that the spinor should transform under general coordinate transformations \emph{by itself}, not as \emph{part of a larger object}, such as one including the metric or its conformal part.   When this tacit assumption is denied, one indeed finds that the spinor-metric complex has a transformation law under arbitrary coordinate transformations, at least sufficiently close to the identity or without any restriction if the metric is positive-definite.  The transformation rule for the spinor is thus metric-dependent and hence nonlinear, though it is linear in the spinor.  This composite object leads to additional terms in Lie differentiation, the neglect of which explains the claim \cite{BennTucker,PenroseRindler2} that spinors only have a coordinate transformation rule for conformal Killing vectors.  The nonlinear group realization formalism gives a Lie derivative of the spinor-(conformal part of) metric complex along any vector field, as will appear below.  Thus it is possible to include spinors as such in coordinates after all, albeit with some technical complication.  This treatment contrasts with treating spinors as coordinate scalars and as spinors with respect to a new internal Lorentz group that acts on an orthonormal basis.  Indeed a common back door to the nonlinear group realization formalism is the imposition of the very popular symmetric tetrad gauge condition in the context of the Einstein-Dirac equation (GR + spin $\frac{1}{2}$) or supergravity (GR + spin $\frac{3}{2}$)  (extensively cited in (\cite{PittsSpinor}). 

 It turns out that the nonlinear group realization formalism was largely invented by Bryce Seligman DeWitt in his dissertation \cite{DeWittDissertation}.  A related work was submitted to the \emph{Physical Review} but, at 85 pages in typescript, was rejected.  This work is now available online \cite{DeWittSpinor1950}.  DeWitt seems not to have grasped the depth of his accomplishment, however.  In the brief DeWitt \& DeWitt paper \cite{DeWittSpinor} the formalism has shrunk to a mere footnote (no. 7) explaining how to use the binomial series to take a symmetric square root of the metric.  In later years DeWitt would echo the conventional Weylian wisdom as though he hadn't constructively refuted it 15 years earlier \cite[p. 115]{DeWittDToGaF}. But inventing nonlinear group realizations is a great accomplishment whether or not one fully realizes or remembers that one has done it, especially as a graduate student and in 1949.  Likely it is  not coincidental that DeWitt/Seligman and Ogievetsky and Polubarinov both used a perturbative expansion and both even used $x^4=ict$, though the latter is not essential and introduces some limitations that  have been noticed more recently \cite{BilyalovSpinors,PittsSpinor,DeffayetSymmetricTetrad}. %
One does not normally expect to make fundamental  conceptual innovations using a binomial series expansion, but this is an example \cite{TenerifeProgressGravity}.  The  symmetric square root of the metric  can appear in the Dirac equation and thus couple gravity to spinor fields.\footnote{I thank Alex Blum for help with the dissertation and the now-available preprint and Dean Rickles for insights into the refereeing process leading to the paper's rejection.}

This paper, then, aims to think foundationally about spinors, to include freedom to change time coordinates and hence more of a spatio-temporal view point than much work in Hamiltonian GR, and to avoid gratuitous gauge symmetry from the tetrad by making use of nonlinear group realizations, thus including all the physically significant gauge freedom (including changes of time coordinates) and no physically insignificant gauge freedom (thus excluding/fixing every part of the tetrad that carries more information than the metric).  Given a $3+1$ split (which fits nicely with Dirac's achievement in trivializing the primary constraints in GR \cite{DiracHamGR}), it is natural to take not the symmetric square root of the space-time metric as many have \cite{OPspinor,PittsSpinor}, but rather the Schwinger time gauge (in effect, locking the time-like tetrad leg to match the unit normal vector defined by the space-time metric and the time coordinate) and the Kibble-Deser symmetric square root of the \emph{spatial} metric \cite{KibbleSymmetric,DeserMoller,DeserCargese} along with the lapse function $N$ and the shift vector $\beta^i$ (which roughly correspond to the time-time and time-space components of the metric \cite[Ch. 21]{MTW}).  Thus one gets a somewhat different nonlinear realization of the space-time coordinate freedom, because (if you like) one is effectively imposing a $3+1$-friendly rather than Lorentz-covariant condition on the tetrad.  If the time-space components of the metric (the ADM shift vector)  disappear from the formalism altogether, as happens in the spatially homogeneous toy theory studied here, the distinction between the two nonlinear formalisms disappears.  


\subsection{Relation to Some Other Projects in Quantum Gravity}

Given that change really does exist in GR, there are many ways to find it. This work being a part of the Lagrangian-equivalent reforming literature (much of which is cited above), it is not necessary to discuss connections to the rest of that literature here.  I have previously discussed in some detail the relations between this project and various other projects in both the physics and philosophy literatures \cite{GRChangeNoKilling}.  
I summarize a few parts of that discussion here. It would be interesting to combine the approach to conditional probabilities taken by Gambini \emph{et al.} \cite{GambiniPullin} with the idea that observables are covariant under coordinate transformations rather than invariant under first-class transformations.

 Rovelli's advice to  ``Forget Time'' \cite{RovelliForgetTime} is, at least in part,
an exhortation to remember space-time and to maintain equivalence to Lagrangian GR (Einstein's equations) in the face of entrenched habits of violating it.  Since \emph{c.} 1960, much work in canonical quantum gravity has made little effort to retain a space-time viewpoint or exhibit mathematically the freedom to change space-time coordinates \cite{BergmannObservables}, a freedom so striking and characteristic of GR. Instead there has been heavy reliance on a space-over-time split to the point of largely discarding space-time in favor of mere space.  If forgetting time means remembering space-time and devising a mathematical formalism that implements it---which requires, for example, using not phase space but phase space-time to accommodate velocity-dependent gauge transformations  \cite{MukundaSamuelConstrainedGeometric,SuganoGeneratorQM,SuganoGaugeGenerator,LusannaVelocityHamiltonian}---then Hamiltonian-Lagrangian equivalence and Rovelli's project overlap significantly.  Rovelli's  ``partial observables''   bear a strong  resemblance to what observables always should have been in Hamiltonian GR, and what they implicitly were in Lagrangian GR, covariant under coordinate transformations (tensors, \emph{etc.}) rather than invariant \cite{ObservablesEquivalentCQG}.  Rovelli's physical/relational point individuation, as opposed to the primitive individuation introduced in modern-style differential geometry \cite{RovelliObservable,RovelliPartialObservables}, is also implemented by using passive coordinate transformations rather than active diffeomorphisms.\footnote{Weatherall's recent work on the hole argument also finds that passive coordinate transformations better reflect reality than do active diffeomorphisms \cite{WeatherallHole}. }

Kuch\v{a}r's doubts that observables need to commute with the Hamiltonian constraint   \cite{KucharCanonical93} and introduction of common-sensical standards motivated rethinking the role of all $4$ secondary constraints in relation to gauge transformations \cite{GRChangeNoKilling}.  
Barbour and Foster's doubts that primary first-class constraints typically generate gauge transformations are also congenial  \cite{BarbourFosterPrimary,FirstClassNotGaugeEM,GRChangeNoKilling}.  I have expressed caution about eliminating the lapse function $N$ from the action as Barbour has proposed \cite{BarbourTimeless1,BarbourTimeless2}, however; such a technique is not available approximately in approximations of GR.  

Related ideas  contribute to E. Anderson's masterful work taking relationism as a premise and finding change \cite{AndersonProblemofTime}.  Anderson takes there to be a problem of time already in classical GR, however, whereas I take the problem to arise (after unjustified postulates have been excluded) in quantization.

Given the  velocity-dependent character of foliation-changing coordinate transformations, one can agree with Th\'{e}bault that it is not very obvious what it would be  be to construct a reduced phase \emph{space} for GR \cite{ThebaultCanonicalGRTime} and that traditional descriptions \cite{BelotEarman} merit reconsideration.\footnote{For a detailed discussion of mathematical presuppositions in Belot's account \cite{BelotPOP}, the reader can see the earlier discussion \cite{GRChangeNoKilling}. 
}  Gryb and Th\'{e}bault find that ``time remains'' and propose an alternative means of quantization which gives the Hamiltonian constraint a distinctive role \cite{GrybThebault,GrybThebaultSchrodinger}.

\subsection{Lagrangian, Notations and Conventions}

The current paper aims to understand the Lagrangian-equivalent approach to Hamiltonian GR with a case rarely considered, GR with spinors, while making full contact with the under-recognized nonlinear `group' realization formalism for spinors in curved space-time.  
Many of the notations and conventions follow Nelson and Teitelboim \cite{NelsonTeitelboim}, including $-+++$ signature, and Majorana spinors with real Dirac numerical $\gamma$ matrices with anticommutation relations $\{ \gamma_A, \gamma_B \} = 2 \eta_{AB}.$ Consequently $\bar{\psi}= \psi^{\dagger} \gamma_0$ and $\gamma_0=-\gamma^0. $
Taking the real and imaginary parts of a complex $4$-spinor $\psi$  yields two real spinors $\phi$ (the real part of $\psi$) and $\chi$ (the imaginary part).
 Some ways that this paper differs from (\cite{NelsonTeitelboim}) include `fixing the tetrad gauge' as soon as possible (because  the tetrad was never really there in the first place given the nonlinear realization) and employing a spatially densitized spinor as soon as possible.  I also avoid notions of an extended Hamiltonian and the idea that individual first-class constraints typically generate gauge transformations, ideas that violate Hamiltonian-Lagrangian equivalence \cite{FirstClassNotGaugeEM,GRChangeNoKilling}.  A sign difference will also appear in the definition of the spin connection.  Some other relevant works include  (\cite{HenneauxGeometrodynamicsTetrad,HenneauxGeneiauEinsteinDirac,HenneauxBianchiIspinor,PonsReduce,PonsSalisburyShepleyAshtekar,SalisburyBianchiI}).  

Some further conventions involve the tetrad and its inverse.  I write the tetrad $e^{\mu}_A$  such that $g^{\mu\nu} = e^{\mu}_A \eta^{AB} e^{\nu}_B.$  The cotetrad $f_{\mu}^A$ is the inverse of $e^{\mu}_A$ on either kind of index and satisfies $g_{\mu\nu} = f_{\mu}^A \eta^{AB} f_{\nu}^B.$   
Greek are world indices, $0$ to $3$.  Capital Latin indices $A, B, M, N\ldots$ are local Lorentz indices running from $0$ to $3$.  Lower-case Latin indices run from $0$ to $3$ and can be the spatial part of either Greek or Latin indices.  Given that the tetrad gauge conditions destroy the distinction between world and local Lorentz indices \cite{KibbleSymmetric} and that the $-+++$ signature implies that spatial indices suffer no sign changes when moved, there should be no confusion.  Indices are rarely moved, but on occasion Greek indices are moved with the space-time metric $g_{\mu\nu}$ or its inverse, local Lorentz indices are more often moved with the Minkowski matrix $diag(-1,1,1,1).$ Occasionally lower-case Latin indices $a, m\ldots$ derived from upper-case Latin indices $A, M\ldots$ are moved with the spatial Kronecker $\delta^i_j$ when there is no ambiguity.  For the ADM shift vector $\beta^n$ relating constant coordinate location with orthogonality to the time hypersurface, indices are moved with the spatial metric $h_{ij} = g_{ij}.$ The inverse spatial metric $h^{\ij}$ is not equal to the spatial part of the inverse space-time metric \cite[ch. 21]{MTW} due to time-space cross-terms involving the shift vector, but the shift vector disappears from the spatially homogeneous case considered here.  
The role of the tetrad is largely heuristic:  the formalism is familiar and makes coordinate covariance manifest, whereas nonlinear group realization formalisms, whether Lorentz-covariant or $3+1$-friendly, are still not so familiar and make coordinate covariance not so obvious. One should not take the initial use of the tetrad formalism as a guide to what actually exists according to the theory in question or  to what topological restrictions are implied.

Using a geometrically rather than physically motivated normalization for the Lagrangian, I take the Lagrangian density to be \cite{NelsonTeitelboim} 
\begin{equation}  \mathcal{L} =\sqrt{-g}R + div + \frac{i}{2}\sqrt{-g} (e^{\mu}_A \bar{\psi} \gamma^A D_{\mu} \psi - (D_{\mu} \bar{\psi}) \gamma^A e^{\mu}_A \psi)   - i m \sqrt{-g} \bar{\psi} \psi. \end{equation} 
The operator $D_{\mu}$ takes the local Lorentz-covariant derivative of the spinor
\begin{eqnarray}    D_{\mu} \psi = \psi,_{\mu} + B_{\mu} \psi, \nonumber \\  
 D_{\mu} \bar{\psi} = \bar{\psi},_{\mu} - \bar{\psi} B_{\mu},  \end{eqnarray}  
where 
\begin{equation} B_{\mu} = \frac{1}{8} e^{\nu}_C \eta^{AC} (f^B_{\nu},_{\mu} - \Gamma_{\nu\mu}^{\alpha} f_{\alpha}^B) [\gamma_A, \gamma_B]. \end{equation} 
While I follow many conventions of Nelson and Teitelboim, this definition uses the opposite sign for the coefficient of $B_{\mu}$ in the covariant derivatives; thus I find that all undesirable terms indeed cancel.  
Here  $\Gamma_{\nu\mu}^{\alpha}$  is the usual Levi-Civita connection (Christoffel symbols) of General Relativity and $[\gamma_A, \gamma_B]$ is the matrix commutator $\gamma_A  \gamma_B  - \gamma_B \gamma_A,$ not to be confused with the strength-$1$ antisymmetric part of the product, $\gamma_{[A} \gamma_{B]} = \frac{1}{2} (\gamma_A \gamma_B - \gamma_B \gamma_A).$  
Notwithstanding the factors of $i$, this Lagrangian is real, thanks to the services of $ \gamma_0.$  $B_{\mu}$ is proportional to the Levi-Civita covariant derivative (giving no attention  to local Lorentz freedom) of the cotetrad $f_{\mu}^A,$ which I write as $f_{\mu}^A;_{\nu}.$ That result makes sense if one envisages fixing the spin connection by requiring  a combined Levi-Civita+spin covariant derivative to annihilate the (co)tetrad.    The relation of the signs of a spinor transformation $\psi^{\prime} = D(\Lambda) \psi$ to the transformation of the tetrad indices follows from requiring the covariance of the Dirac equation, yielding  $D(\Lambda)^{-1} \gamma^N D(\Lambda) = \gamma^P \Lambda^N_{.P}.$  The cotetrad $f_{\mu}^A,$ a collection of covariant vector fields under coordinate transformations of the index $\mu,$ is a set of \emph{contravariant} vector fields under local Lorentz transformations:  $f^A_{\mu} \rightarrow f^A_{\mu} + \Omega^A_{.B} f^B_{\mu},$ where $\Omega^A_{.B}$ is infinitesimal and is antisymmetric when an index is moved with $diag(-1,1,1,1).$  Likewise the tetrad, composed of contravariant world vectors, is composed of local Lorentz covectors:  $e^{\mu}_A \rightarrow e^{\mu}_A - \Omega^B_{.A}e^{\mu}_B.$  Then $\psi \rightarrow (I + \frac{1}{4}\gamma_{[A} \gamma_{B]})\psi.$

One  step in simplifying the Lagrangian is the imposition of the Schwinger time gauge \cite{KibbleSymmetric}, which locks the time-like leg of the tetrad $e^{\mu}_0$ to match the future-pointing unit normal vector $n^{\mu}$ defined by the space-time metric and the time coordinate's gradient.  Thus the time-like leg of the cotetrad $f^0_{\mu}$ matches the (negative) covariant unit normal vector $-n_{\mu}$, which in terms of the ADM lapse function $N$ and shift vector $\beta$ satisfies $n^{\mu}= (\frac{1}{N}, -\frac{\beta^m}{N})$  \cite[p. 508]{MTW}.    
  While the covariant unit normal $n_{\mu}$ is closer to fundamental than the contravariant one $n^{\mu}$ due to the former's using the metric only to rescale the gradient of the time coordinate, not also moving an index, a minus sign has to go somewhere (given $-+++$ signature, which choice is crucial once the distinction between covariant and contravariant spatial indices disappears below). The fact that contravariant vectors more straightforwardly point in a direction is perhaps sufficient grounds for setting $e^{\mu}_0 = n^{\mu}$ with no sign-flip.  The covector unit normal, in terms of the lapse function $N$, is $n_{\mu} =(-N,0,0,0).$   The relation $f^0_{\mu}= -n_{\mu}$ implies $f^0_m=0$ \cite[p. 508]{MTW}:  the spatial components  of the time-like co-leg vanish.  The orthogonality of the space-like triad of legs $e^{\mu}_a$ to the normal covector $n_{\mu}$ implies $e^0_a=0$:  the time components of the spatial triad vanish.  One can also show that $f^a_0 f^a_n = \beta_n,$   $f^0_0=N,$ and $f^a_m f^a_n = h_{mn}.$  Quite apart from talk of tetrads one has the relation between spatio-temporal and spatial volume elements $\sqrt{-g} = N \sqrt{h}.$

Another step  that I will take is a fairly naive imposition of spatial homogeneity by simply dropping all spatial derivatives \cite{GRChangeNoKilling}.  This is, of course, both a strong restriction of the physics and a substantial fixation of the spatial coordinate freedom.  (For more careful treatments of the spatial coordinate freedom, see \cite{HenneauxBianchiIspinor,AshtekarBianchi,PonsReduce}.)
My interest is in doing justice to the temporal coordinate freedom, something rarely attempted and best accomplished by keeping the spatial metric look as much like it does in full GR as possible, rather than keeping track of the remnant of spatial coordinate freedom that one would have by interpreting the Lagrangian as derived from GR.  One can assume instead that one is simply handed the Lagrangian that results from the naive spatial truncation. This procedure will indicate more spatial degrees of freedom than one should inherit from GR, but that is not important for the task at hand.  One could impose the initial vanishing of such extra degrees of freedom to recover a closer match to GR in that respect if one wished; such a condition would be dynamically preserved.

After expanding out the covariant derivative of the spinors into partial derivative and connection terms, one can give the spinors  a \emph{spatial} density weight of $\frac{1}{2}$  \cite{KibbleSymmetric,NelsonTeitelboim,HenneauxBianchiIspinor}, thereby removing some interaction terms between the spinor and gravity.  This densitization will be denoted by a $\sim$ over spinors $\psi,$  $\bar{\psi},$ $\phi,$ $\chi,$ \emph{etc}. (Note that one cannot readily write half a $\sim$; the notion of expressing the weight of a density by the number of tildes was never going to be viable for densities other than $\pm1$ and maybe $\pm2$; fractional and irrational weights, though perfectly well defined \cite{Schouten,Anderson} and sometimes useful \cite{OPspinor,MassiveGravity1,MassiveGravity2,MassiveGravity3,PittsScalar,PittsSpinor}, were never going to be accommodated in detail typographically.)  
 The Lagrangian density then takes the form 
\begin{eqnarray} \mathcal{L}= \sqrt{-g}R + div - 2mN{\stackrel{\sim}{\chi}^{\top}}\gamma_0\stackrel{\sim}{\phi}   -{ Ne^{\mu}_A\stackrel{\sim}{\chi},_{\mu}^{\top}\gamma_0\gamma^A\stackrel{\sim}{\phi} } +  Ne^{\mu}_A\stackrel{\sim}{\chi}^{\top}\gamma_0\gamma^A \stackrel{\sim}{\phi},_{\mu} \nonumber \\   + 2 N e^{\mu}_A \stackrel{\sim}{\chi}^{\top} \gamma_0 B_{\mu} \gamma^A \stackrel{\sim}{\phi} + N  \stackrel{\sim}{\chi}^{\top} \gamma_0 \gamma_B g^{\mu\nu} f^B_{\nu;\mu} \stackrel{\sim}{\phi}.
  \end{eqnarray}

One can define canonical momenta for the real spinors with the usual definition: 
\begin{eqnarray} 
\pi_{\phi} = \frac{\partial \mathcal{L} }{\partial \dot{\stackrel{\sim}{\phi}} } = N e^{\mu}_A \stackrel{\sim}{\chi}^{\top} \gamma_0 \gamma^A \delta^0_{\mu} = N e^{0}_A \stackrel{\sim}{\chi}^{\top} \gamma_0 \gamma^A, \nonumber \\
\pi_{\chi} = \frac{\partial \mathcal{L} }{\partial \dot{\stackrel{\sim}{\chi}} } = -N e^{\mu}_A  \gamma_0 \gamma^A \delta^0_{\mu}\stackrel{\sim}{\phi} = -N e^{0}_A \gamma_0 \gamma^A \stackrel{\sim}{\phi}. 
\end{eqnarray}
Imposing the Schwinger time gauge gives a considerable simplification because the time components of the spatial legs vanish: $e^0_a=0$.  One also has $e^0_0=\frac{1}{N}.$  Using the relation $\gamma_0 \gamma_0=-I,$ the canonical momenta simplify to 
\begin{eqnarray} 
\pi_{\phi} =  \stackrel{\sim}{\chi}^{\top},  \\
\pi_{\chi} =  -  \stackrel{\sim}{\phi}. 
\end{eqnarray}
Clearly none of these relations can be solved for the velocities.  Hence we have $8$ primary constraints $\pi_{\phi}-  \stackrel{\sim}{\chi}^{\top}, $    $ \pi_{\chi} +   \stackrel{\sim}{\chi}$, quantities that involve canonical momenta and vanish due to the impossibility of the Legendre transformation from velocities to  momenta.


\section{Constrained Dynamics of Spinors without Gravity} 

The assumptions made so far (the time gauge and the spatial weight $\frac{1}{2}$ definition of the spinors) have led to a simple Lagrangian for the spinor fields, but spinor fields behave in a sufficiently unfamiliar fashion in constrained Hamiltonian dynamics to justify forgetting about gravity altogether for the moment.  Forgetting about gravity can be effected by setting (temporarily) $f_{\mu}^A $ to the Kronecker delta (identity matrix).  
The canonical Hamiltonian is defined with the usual sum over momenta times velocities, including the constrained quantities, but then the primary constraints are used to try to simplify the result \cite{Sundermeyer}.
In contrast to cases where the Lagrangian is quadratic in a velocity (permitting the replacement of the velocity by a momentum) or independent of a velocity (as occurs for the electrostatic scalar potential in electromagnetism and for the lapse function and shift vector in GR and so yielding vanishing canonical momenta), the Lagrangian here is linear in the velocity.  Thus the momenta are nonzero but independent of the velocity in question, the worst of both worlds.  Pressing on, one finds that
\begin{eqnarray} \mathcal{H}_c = (\pi_{\phi} \dot{\stackrel{\sim}{\phi}}  +   \pi_{\chi} \dot{\stackrel{\sim}{\chi}} - \mathcal{L} )|_{primaries} \nonumber \\
= 2 m \chi^{\top} \gamma_0 \phi + \chi^{\top},_i \gamma_0 \gamma^i \phi - \chi^{\top} \gamma_0 \gamma^i \phi,_i.
\end{eqnarray} 
The densitization has been dropped with the neglect of gravity and consequent specialization to Cartesian coordinates, so no $\sim$ symbols are needed.  Note that the momenta are absent from the canonical Hamiltonian.  
The quantity that gives Lagrangian-equivalent Hamilton's equations, however, is the total Hamiltonian \cite{Sundermeyer}, which includes terms involving the primary constraints multiplied by either the corresponding velocities 
 or some arbitrary functions that turn out \emph{a posteriori} to equal the velocities in question: 
\begin{eqnarray} 
\mathcal{H}_p = \mathcal{H}_c + (\pi_{\phi} - \chi^{\top}) \dot{\phi}  + \dot{\chi}^{\top} (\pi_{\chi} + \phi).  
\end{eqnarray} 
The Hamiltonian equations for $\phi$ are \begin{eqnarray} \dot{\phi} = \{\phi, \mathcal{H}_p \} = \{ \phi, \mathcal{H}_c \} + \dot{\phi} + 0 = \dot{\phi}, \end{eqnarray}
which is true, though not informative, and
\begin{eqnarray} \dot{\pi_{\phi} } = \{ \pi_{\phi}, \mathcal{H}_p \} = - \left( \frac{\partial \mathcal{H}_c }{\partial \phi } - \frac{\partial}{\partial x^i} \frac{\partial  \mathcal{H}_c }{\partial \phi,_i } \right)  +  \{ \pi_{\phi} , (\pi_{\phi} - \chi^{\top}) \dot{\phi} \} + \dot{\chi}^{\top} \{ \pi_{\phi}, \pi_{\chi} + \phi \} \nonumber \\
 = -2 \chi^{\top},_i \gamma_0 \gamma^i -2m\chi^{\top} \gamma_0 - \dot{\chi}^{\top}. 
\end{eqnarray} 
We did not need to commit much to the meaning of a Poisson bracket of a velocity because every such term  was multiplied by a primary constraint.   
If we allow ourselves already  to use the time derivative of a primary constraint, then this Hamiltonian equation is seen to be equivalent to the Dirac equation for $\chi^{\top}$ (or $\bar{\chi}$)---not for $\phi.$ 
Hamilton's equations for $\chi$ are analogous:  $\{ \chi^{\top}, \mathcal{H}_p \}$ is the empty relation $\dot{\chi}^{\top} = \dot{\chi}^{\top},$ while $\{ \pi_{\chi}, \mathcal{H}_p \}$ gives a relation that (adding in the time derivative of a primary constraint) is the Dirac equation for $\phi.$

In the interest of distinguishing first-class constraints (which relate to gauge freedom) from second-class constraints (which do not), a process that can require taking linear combinations of constraints in order to find the appropriate number of first-class constraints, one wants to ensure that the primary constraints, if true initially, are preserved by the dynamics.  
Preservation of $\pi_{\phi} - \chi^{\top}$'s vanishing  yields the Dirac equation for $\chi^{\top},$ which, containing $\dot{\chi}^{\top},$ is not a constraint.  Analogously, preserving $\pi_{\chi} + \phi$'s vanishing yields the Dirac equation for $\phi,$ which is not a constraint.  Thus there are no secondary constraints for the pure spinor system without gravity.  The previously undetermined velocities are, however, fixed in terms of phase space quantities, as one expects for second-class constraints.  If one wants the $8 \times 8$ matrix (at each spatial point) of the Poisson brackets of all the constraints among themselves, the nonzero contributions come from the two copies of the $4 \times 4$ relation $$ \{ \pi_{\phi}^{\top}(x) - \chi(x),  \pi_{\chi}(y) + \phi^{\top}(y) \} = -2 I \delta(x,y).$$


\section{Gravity and Spinors without Spatial Dependence}

Having now an idea of how the spinor matter behaves, let us restore gravity but discard all spatial derivatives in order to focus attention on temporal coordinate freedom and change with as little technical complication as possible.  
The Schwinger time gauge is employed.  The Lagrangian can be taken as above.  Discarding some obviously vanishing terms such as the spatial Ricci scalar but taking more time to pare down the partial derivative of the cotetrad (which will be the source of the surviving term involving the rotation in spinor space) and Christoffel symbol term (which cancels some undesirable terms from the partial derivative of the cotetrad), making use of the anticommutation relations of the $\gamma$ matrices and the usual tricks of tensor calculus, 
after a page or two of algebra one obtains
\begin{eqnarray} 
\mathcal{L} = N \sqrt{h}(K_{ij} K^{ij} - K^2) - 2 m N \stackrel{\sim}{\chi}^{\top} \gamma_0 \stackrel{\sim}{\phi} - \stackrel{\sim}{\chi},_0^{\top} \stackrel{\sim}{\phi} +  \stackrel{\sim}{\chi}^{\top} \stackrel{\sim}{\phi},_0 +  \frac{1}{2} \stackrel{\sim}{\chi}^{\top} e^n_a f_{nb},_0 \gamma_{[a} \gamma_{b]}\stackrel{\sim}{\phi}.   
\end{eqnarray} 
Reassuringly, this expression does not contain $\dot{N}$ or $\dot{\beta}^i$ and contains $N$ and $\beta^i$ at most linearly, as one expected from the coupling of more familiar matter in GR. Given the spatial homogeneity, one is not surprised that $\beta^i$ is absent.  
 The last term, which involves  a rotation in spinor space, seems not to be removable using a field redefinition.  It does, however, depend merely on the conformal part $\hat{h}_{ij}$ of the spatial metric $h_{ij}$. One could see that, for example, by factoring the triad (and inversely the cotriad) into a unimodular part (unit determinant) and the appropriate power of the volume element and then notice how the volume element terms yield a $\delta_{ab}$ that cancels when contracted with the antisymmetric term in the $\gamma$ matrices.

One straightforwardly verifies that this Lagrangian is invariant under `local' (that is, time-dependent) rotations: the original local Lorentz invariance described by  $\Omega^A_{.B}(t,x,y,z)$ has been  whittled down to $\Omega^a_{.b}(t),$ under which the Lagrangian is invariant.    The period serves as a placeholder to distinguish the indices of $\Omega^A_{.B}(t,x,y,z),$ which is antisymmetric when the indices are at the same level.

The presence of the velocity of the spatial metric implies that the canonical momenta for gravity are altered by the presence of spinors.  The use of the Schwinger time gauge and the weight $\frac{1}{2}$ densitization gets rid of many inconvenient terms \cite{NelsonTeitelboim}.  The Kibble-Deser symmetric triad condition gives further leanness. Given that my goal ultimately is to understand time coordinate freedom in GR, not in this toy theory, it would be unhelpful to try to remove $3$ degrees of freedom per spatial point using the remnant of the spatial coordinate freedom.  
  At this point plausibly all the fat has been removed, leaving only muscle and bone (unless one is prepared to seek the unconstrained true degrees of freedom at the expense of locality).  So plausibly this use of gauge conditions and field redefinitions has put the  Lagrangian in as convenient a form as possible.  
One has to decide what variables to use as canonical coordinates.  An attractive choice native to spinors would be to use the symmetric cotriad $f_{ai}$ or possibly its inverse $e^{ai}.$  While such a treatment would be a worthwhile project, certain inversions would become complicated.  In the interest of familiarity, let us use the usual geometrodynamic quantity $h_{ij}=g_{ij},$ the spatial metric tensor.  This choice will leave some square roots that are not so readily simplified, but they will not cause trouble.

While the canonical momenta for the spinor fields are as given above in the absence of gravity, the canonical momenta for gravity undergo some modification.  While $\pi_N = \frac{\partial \mathcal{L} }{ \partial \dot{N} } = 0$ as usual in GR, the spatial metric's canonical momenta acquire a contribution from the spinor fields. 
\begin{eqnarray} 
\pi^{ij} = \frac{\partial \mathcal{L}_{GR} }{\partial \dot{h}_{ij} } + \frac{\partial}{\partial \dot{h}_{ij} } \frac{ \stackrel{\sim}{\chi^{\top} }}{2} \sqrt{  \hat{h}^{na} } \sqrt{ \hat{h}_{nb} },_0 \gamma_{ [a} \gamma_{b]} \stackrel{\sim}{\phi}.
\end{eqnarray} 
The trace $\pi = \pi^{ij} h_{ij}$ of these canonical momenta receives no contribution from the spinor term, as one can see either from the presence of only the conformal part of the metric $\hat{h}_{ij}$ or from the antisymmetrization of the indices of the $\gamma$ matrices. Solving for the velocities of the spatial metric gives 
\begin{eqnarray} 
\dot{h}_{cd} = \frac{2N}{\sqrt{h} }  (\pi_{cd} - \frac{ h_{cd} }{2} \pi - h_{ic} h_{dj}  \frac{ \tilde{\chi} }{2} \sqrt{h^{na}} \frac{ \partial \sqrt{h_{nb}} }{ \partial h_{ij} } \gamma_{[a} \gamma_{b]} \stackrel{\sim}{\phi}).
\end{eqnarray} 

 The canonical Hamiltonian is \begin{eqnarray} 
\mathcal{H}_c = N \pi_N + \pi^{cd} \dot{h}_{cd} + \pi_{\phi} \dot{ \stackrel{\sim}{\phi}} + \dot{ \stackrel{\sim}{\chi} }^{\top}  \pi_{\chi} - \mathcal{L}  \nonumber \\
= 0 + \pi^{cd} \cdot  \frac{2N}{\sqrt{h} }  (\pi_{cd} - \frac{ h_{cd} }{2} \pi - h_{ic} h_{dj}  \frac{ \tilde{\chi} }{2} \sqrt{h^{na}} \frac{ \partial \sqrt{h_{nb}} }{ \partial h_{ij} } \gamma_{[a} \gamma_{b]} \stackrel{\sim}{\phi}) 
+ \stackrel{\sim}{\chi}^{\top}  \dot{ \stackrel{\sim}{\phi} }  + \dot{ \stackrel{\sim}{\chi}}^{\top} \cdot -\stackrel{\sim}{\phi}   \nonumber \\ 
- (N \sqrt{h} K^{ij}K_{ij} - N \sqrt{h} K^2 - 2 m N \stackrel{\sim}{\chi}^{\top} \gamma_0 \stackrel{\sim}{\phi} - \stackrel{\sim}{\chi}^{\top},_0 \stackrel{\sim}{\phi} + \stackrel{\sim}{\chi} \stackrel{\sim}{\phi},_0 + 
 \frac{ \stackrel{\sim}{\chi}^{\top} }{2} \sqrt{h^{na}} \sqrt{h_{nb} },_0 \gamma_{[a} \gamma_{b]} \stackrel{\sim}{\phi} ) \nonumber  \\ 
= \frac{N}{\sqrt{h} } \pi^{cd} \pi_{cd} -\frac{N}{2 \sqrt{h} } \pi^2  - \frac{N}{\sqrt{h} } \pi_{ij} \frac{ \stackrel{\sim}{\chi}^{\top} }{2} \sqrt{h^{na} } \frac{ \partial \sqrt{ h_{nb}} }{\partial h_{ij} } \gamma_{[a} \gamma_{b]} \stackrel{\sim}{\phi}  %
+ 2 m N \stackrel{\sim}{\chi}^{\top} \gamma_0 \stackrel{\sim}{\phi} +   \nonumber \\
 \frac{N}{2 \sqrt{h} } \stackrel{ \sim}{\chi}^{\top} \gamma_{[a} \gamma_{b]} \stackrel{\sim}{\phi} \frac{ \stackrel{\sim}{\chi}^{\top} }{2} \gamma_{[e} \gamma_{f]}  \stackrel{\sim}{\phi} h_{ic} h_{jd} \sqrt{h^{na} } \frac{ \partial \sqrt{h_{nb} }}{\partial h_{ij} } \sqrt{h^{me} } \frac{ \partial \sqrt{h_{mf} }}{\partial h_{cd} } = N \mathcal{H}_0,
\end{eqnarray}
where the last relation is a definition of what will turn out to be the Hamiltonian constraint; I have felt free to include the matter contribution within the symbol $\mathcal{H}_0$ for brevity because I will not make separate use of the gravitational and material pieces.  
Clearly this system is reparametrization-invariant because the entire expression involves the lapse linearly.  Hence as much temporal general covariance as a spatially homogeneous system could have is evident.

The primary Hamiltonian \cite{Sundermeyer} adds to the canonical Hamiltonian terms involving the primary constraints multiplying the corresponding velocities so as to yield Hamilton's equations that are mathematically equivalent to the Euler-Lagrange equations. 
 One has
\begin{eqnarray} 
\mathcal{H}_p = N \mathcal{H}_0 + \pi_N \dot{N} + (\pi_{\phi} - \stackrel{\sim}{\chi}^{\top} ) \dot{\stackrel{\sim}{\phi} } + \dot{ \stackrel{\sim}{\chi}}^{\top} (\pi_{\chi} + \stackrel{\sim}{\phi}). 
\end{eqnarray}

This system has $9$ primary constraints, including $\pi_N$ inherited from GR and $8$ primary constraints from the spinors.  The dynamics needs to preserve the primary constraints.  For $\pi_N$ one has 
\begin{eqnarray} 
\{\pi_N, \mathcal{H}_p \} = \{ \pi_N,  N \mathcal{H}_0 + \pi_N \dot{N} + (\pi_{\phi} - \stackrel{\sim}{\chi}^{\top} ) \dot{\stackrel{\sim}{\phi} } + \dot{ \stackrel{\sim}{\chi}}^{\top} (\pi_{\chi} + \stackrel{\sim}{\phi} \}  = - \mathcal{H}_0 \stackrel{!}{=} 0,
\end{eqnarray} 
showing that (minus) $\mathcal{H}_0$ is indeed a constraint.  
The term $\pi_N \dot{N}$ yields nothing using the primary constraint or employing the sometimes-needed Anderson-Bergmann velocity Poisson bracket \cite{AndersonBergmann,ObservablesEquivalentCQG}.
Preservation of the primary constraint $\pi_{\phi} - \stackrel{\sim}{\chi}^{\top} $  yields 
\begin{eqnarray} 
\{ \pi_{\phi} - \stackrel{\sim}{\chi}^{\top}, \mathcal{H}_p \} = - N \frac{\partial \mathcal{H}_0 }{\partial \stackrel{\sim}{\phi} } - 2 \dot{\stackrel{\sim}{\chi}}^{\top} \stackrel{!}{=}0.
\end{eqnarray} 
This equation involves a velocity and so, rather than yielding a new constraint, fixes the velocity for $\stackrel{\sim}{\chi},$ as one expects given the second-class character of the constraint in the absence of gravity.  
Apart from certain signs, the same story applies for $\pi_{\chi} + \stackrel{\sim}{\phi}:$ 
\begin{eqnarray} 
\{ \pi_{\chi} + \stackrel{\sim}{\phi}, \mathcal{H}_p \} = - N \frac{ \partial \mathcal{H}_0 }{\partial \stackrel{\sim}{\chi}^{\top} } + 2 \dot{\stackrel{\sim}{\phi}}  \stackrel{!}{=}0.
\end{eqnarray}


\section{First-Class and Second-Class Constraints}

One can take Poisson brackets among each pair of constraints and ascertain which constraints are first-class or second-class.  If one is lucky, then the distinction will be clean using the original constraints.  More generally, one might need to take linear combinations of constraints in order to find the expected number of first-class constraints. This expected number, though strongly suggested by the component theories (here GR and spinors satisfying the Dirac equation), follows from the extent by which the matrix rank (the total number of nonvanishing eigenvalues) of the matrix of Poisson brackets falls short of the dimension of the square matrix in question, in this case $10$.  
One easily sees that $\pi_N$ has vanishing Poisson brackets with all the constraints, hence remaining first-class.  One also sees that the brackets among the spinor constraints are unchanged by the introduction of gravity, though of course the spatial Dirac $\delta(x,y)$ functions disappear when one discards spatial dependence.  The more significant results, however, are 
\begin{eqnarray} 
\{ \pi_{\phi}, \mathcal{H}_0 \} = -\frac{\partial \mathcal{H}_0 }{\partial \stackrel{\sim}{\phi} }, \\
 \{ \pi_{\chi}, \mathcal{H}_0 \} = -\frac{\partial \mathcal{H}_0 }{\partial \stackrel{\sim}{\chi} },  
\end{eqnarray}
so the original Hamiltonian constraint is no longer first-class.  This of course does not mean that the gauge freedom to change time coordinates has disappeared. 
Thus the matrix of Poisson brackets of the constraints, ordered as $\pi_N$, $\pi_{\phi} - \stackrel{\sim}{\chi}^{\top},$ $\pi_{\chi} + \stackrel{\sim}{\phi}^{\top},$ $\mathcal{H}_0,$ is 
\begin{eqnarray} 
\left[
\begin{array}{cccc}
\{ \pi_N, \pi_N \}  &  \{ \pi_N, \pi_{\phi} - \stackrel{\sim}{\chi}^{\top} \}   &     \{ \pi_N, \pi_{\chi} + \stackrel{\sim}{\phi}^{\top} \}   &    \{ \pi_N, \mathcal{H}_0\}   \\ 
 \{ \pi_{\phi}^{\top} - \stackrel{\sim}{\chi}, \pi_N \}  &  \{ \pi_{\phi}^{\top} - \stackrel{\sim}{\chi}, \pi_{\phi} - \stackrel{\sim}{\chi}^{\top} \} & \{ \pi_{\phi}^{\top} - \stackrel{\sim}{\chi}, \pi_{\chi} + \stackrel{\sim}{\phi}^{\top} \}  &  \{ \pi_{\phi}^{\top} - \stackrel{\sim}{\chi}, \mathcal{H}_0  \}  \\
\{ \pi_{\chi}^{\top} + \stackrel{\sim}{\phi}, \pi_N \}  &  \{ \pi_{\chi}^{\top} + \stackrel{\sim}{\phi}, \pi_{\chi} + \stackrel{\sim}{\phi}^{\top} \} & \{ \pi_{\chi}^{\top} + \stackrel{\sim}{\phi}, \pi_{\chi} + \stackrel{\sim}{\phi}^{\top} \}  &  \{ \pi_{\phi}^{\top} + \stackrel{\sim}{\chi}, \mathcal{H}_0  \}  \\
\{ \mathcal{H}_0, \pi_N \}  &  \{ \mathcal{H}_0, \pi_{\chi} + \stackrel{\sim}{\phi}^{\top} \} & \{ \mathcal{H}_0, \pi_{\chi} + \stackrel{\sim}{\phi}^{\top} \}  &  \{ \mathcal{H}_0 , \mathcal{H}_0  \}  
\end{array}
\right] = 
\left[
\begin{array}{cccc}
0 & 0_{1\times4} &  0_{1\times4} & 0 \\
0_{4\times1} & 0_{4\times 4} & -2I_{4 \times 4} & - \frac{\partial \mathcal{H}_0^{\top} }{\partial \stackrel{\sim}{\phi} }  \\
0_{4\times1} & 2I_{4\times 4} & 0_{4 \times 4} & - \frac{\partial \mathcal{H}_0^{\top} }{\partial \stackrel{\sim}{\chi} }  \\
0 &   \frac{\partial \mathcal{H}_0^{\top} }{\partial \stackrel{\sim}{\phi} } & \frac{\partial \mathcal{H}_0^{\top} }{\partial \stackrel{\sim}{\chi} }  & 0
\end{array}
\right].
\end{eqnarray} 
The failure of the last column and the last row to vanish displays how the original Hamiltonian constraint is no longer first-class.  
 The facts that $\pi_N$ is first-class and that second-class constraints have to come in even numbers given a finite number of degrees of freedom \cite[p. 80]{Sundermeyer} assure us that there must be some combination of $\mathcal{H}_0$ and the spinor constraints that is first-class.

It is not difficult to find the modified Hamiltonian constraint $\bar{\mathcal{H}}_0$ that, by incorporating contributions from the second-class spinor primary constraints, is first-class.  
An easy way to find it is to devise largely arbitrary combination $\bar{\mathcal{H}}_0 = \mathcal{H}_0 + A (\pi_{\phi}^{\top} - \stackrel{\sim}{\chi} ) + B (\pi_{\chi}^{\top} + \stackrel{\sim}{\phi}),$ where $A$ and $B$ are potentially phase space-dependent row matrices; their Poisson bracket contributions are rendered unimportant due to multiplication by the second-class primary constraints.  Demanding that this  $\bar{\mathcal{H}}_0 $ have vanishing Poisson brackets with both of the spinor constraints implies that 
\begin{eqnarray} 
A= \frac{1}{2} \frac{\partial \mathcal{H}_0 }{\partial \stackrel{\sim}{\chi}  }, \nonumber \\
B = -\frac{1}{2} \frac{\partial \mathcal{H}_0 }{\partial \stackrel{\sim}{\phi}  }.
\end{eqnarray} 
By antisymmetry the bracket with $\bar{\mathcal{H}}_0 $ with itself of course vanishes.  The independence of $\bar{\mathcal{H}}_0 $ from the lapse implies the one remaining vanishing bracket.  
We can now use a redefined set of constraints, 
 $\pi_N$, $\pi_{\phi} - \stackrel{\sim}{\chi}^{\top},$ $\pi_{\chi} + \stackrel{\sim}{\phi}^{\top},$ $\bar{\mathcal{H}}_0,$ thereby making the inconvenient last column and last row vanish.  

One can define a Dirac bracket $\{ , \}^*$ using the $8$ second-class spinor primary constraints using the invertible  $8 \times 8$ matrix of Poisson brackets, which is the same given the use of $\bar{\mathcal{H}}_0$ as for $\mathcal{H}_0$:  
\begin{eqnarray} 
 C_{AB}^8 = 
\left[
\begin{array}{cc}
   \{ \pi_{\phi}^{\top} - \stackrel{\sim}{\chi}, \pi_{\phi} - \stackrel{\sim}{\chi}^{\top} \} & \{ \pi_{\phi}^{\top} - \stackrel{\sim}{\chi}, \pi_{\chi} + \stackrel{\sim}{\phi}^{\top} \}    \\
  \{ \pi_{\chi}^{\top} + \stackrel{\sim}{\phi}, \pi_{\chi} + \stackrel{\sim}{\phi}^{\top} \} & \{ \pi_{\chi}^{\top} + \stackrel{\sim}{\phi}, \pi_{\chi} + \stackrel{\sim}{\phi}^{\top} \} 
\end{array}
\right]   \nonumber \\
 = 
\left[
\begin{array}{cc}
0_{4\times 4} & -2I_{4 \times 4} \\
 2I_{4\times 4} & 0_{4 \times 4}  \\
  \end{array}
\right]
\end{eqnarray} 
with inverse 
  \begin{eqnarray} C^{AB}_8 = 
\left[
\begin{array}{cc}
 0_{4\times 4} & \frac{1}{2}I_{4 \times 4}   \\
 -\frac{1}{2}I_{4\times 4} & 0_{4 \times 4}  
 \end{array}
\right]_.
\end{eqnarray} 
With this Dirac bracket in hand, one can take the second-class spinor constraints $\pi_{\phi}^{\top} - \stackrel{\sim}{\chi}$ and $\pi_{\chi}^{\top} + \stackrel{\sim}{\phi}$ as identities, eliminating $\pi_{\chi} $ and either $\stackrel{\sim}{\chi}$ or alternatively $\pi_{\phi}$ (the two being equal).  As a consequence one finds that $\{ \stackrel{\sim}{\phi}, \pi_{\phi} \}^* = I_{4\times 4} - \frac{1}{2} I_{4\times 4}= \frac{1}{2} I_{4\times 4}$:  these brackets will be only half as large as a standard canonical pair.



\section{Gauge Generator for Changes of Time Coordinate}

Contrary to the common belief that one cannot implement changes in time coordinates and to the simple neglect of the question in some standard GR texts that offer a version of a Hamiltonian formulation \cite[ch. 21]{MTW} \cite[Appendix E]{Wald}, one can implement changes of time coordinate using a Hamiltonian formalism.  One encounters the quirk, harmless at the classical level, that one needs to use Hamilton's equations more often than one might have expected \cite{FradkinVilkoviskyHLEquivalence,CastellaniGaugeGenerator}.  While there are procedures to find the gauge generator by iterating beginning with the primary constraints, another approach\footnote{
I observe that certain minor differences in the treatment of velocities make it appropriate to be somewhat exploratory regarding the gauge generator(s) despite the substantial amount of work (\emph{e.g.}, \cite{MukundaSymmetries,MukundaGaugeGenerator,PonsDirac}) done on the subject.  Mukunda's work, following the book with Sudarshan \cite{SudarshanMukunda}, retains the unsolved-for velocities (here $\dot{N}$) along with the momenta unrestricted by primary constraints in order to get a complete collection of variables with as much information as the velocities in the Lagrangian formalism.  Other authors avoid those velocities in favor of arbitrary functions.  Mukunda's inclusion of velocities (followed by Castellani \cite{CastellaniGaugeGenerator}) does not, however, give velocities any contribution to Poisson brackets, such as the Anderson-Bergmann velocity Poisson bracket does.  Much of that work is implemented/obviated by defining the variation of a velocity to be the velocity of a variation \cite{PonsDirac}, while many other appearances of velocities are multiplied by constraints and thus make no difference in most contexts.  The approach followed in this paper retains the unsolved velocities and uses the Anderson-Bergmann velocity Poisson bracket.  One needs to attend to such matters before carrying over a derivation that makes different choices on this point of fine detail.   
} is simply to add all the constraints together with arbitrary (potentially field-dependent) coefficients and require that the resulting quantity change the canonical action $\int dt d^3x (p \dot{q} - \mathcal{H}_p)$ by at most a boundary term, where $p \dot{q}$ is a `large' sum over all the canonical coordinates (including those related for which the velocities yield primary constraints) and the relevant Hamiltonian is the primary Hamiltonian.
Such a criterion will preserve the Euler-Lagrange equations for the canonical action; because those equations are just Hamilton's equations, Hamilton's equations are preserved.  An advantage of using the canonical action is that it shows that the Hamiltonian formalism is a \emph{special case} of the Lagrangian formalism and hence has neither need nor room for radical conceptual innovations or independent postulates.  
The spatial integral is of course redundant for the spatially homogeneous example at hand.  The temporal integral provides another way to handle, or rather avoid, the Poisson bracket with a velocity.  
The canonical action is thus, after some cancellations, the integral of 
\begin{eqnarray} 
\mathcal{L}_c = \pi^{ij} \dot{h}_{ij} + \stackrel{\sim}{\chi}^{\top}  \dot{\stackrel{\sim}{\phi} }    - \dot{\stackrel{\sim}{\chi}}^{\top} \stackrel{\sim}{\phi}  -N \mathcal{H}_0.
\end{eqnarray}

 It is convenient at this stage to find the Poisson brackets of all 4 sets of constraints $\pi_N$, $\pi_{\phi} -  \stackrel{\sim}{ \chi}^{\top},$  $\pi_{\chi} + \stackrel{\sim}{\phi}^{\top},$ and $\mathcal{H}_0$ (discarding the sign in the last case), having respectively $1,$ $4$, $4$ and $1$ constraint; these are the original constraints rather than the redefined ones, but that should not matter because any special first-class linear combinations ought to reassert themselves.
 Using an arbitrary smearing function $\xi(t),$ one has \begin{eqnarray}  \{ \xi(t) \pi_N, p \dot{q} - \mathcal{H}_p \}  = \xi \mathcal{H}_0. \end{eqnarray} 
Using a column matrix of smearing functions $\epsilon_1(t)$ with $\pi_{\phi} - \stackrel{\sim}{\chi}^{\top},$ one obtains
\begin{eqnarray} \{  (\pi_{\phi} - \stackrel{\sim}{\chi}^{\top}  ) \epsilon_1, p \dot{q}- \mathcal{H}_p  \}   = - \stackrel{\sim}{\chi}^{\top} \dot{\epsilon}_1 + \dot{\stackrel{\sim}{\chi} }^{\top}  \epsilon_1 + N   \frac{\partial \mathcal{H}_0 }{\partial \stackrel{\sim}{\phi} } \epsilon_1
\end{eqnarray} 
using the Anderson-Bergmann velocity Poisson bracket with its surprising property $\{ \dot{q}, F \} = \frac{\partial}{\partial t} \{q, F\}$  \cite{AndersonBergmann}, which supersedes the Poisson bracket product rule when the velocity is isolated \cite{ObservablesEquivalentCQG}.  
Analogously using a row matrix of smearing functions $\epsilon_2(t)$ one has
\begin{eqnarray} \{ \epsilon_2 ( \pi_{\chi}^{\top} + \stackrel{\sim}{\phi} ) ,  p \dot{q}- \mathcal{H}_p \} = - \epsilon_2 \dot{\stackrel{\sim}{\phi} } + \dot{\epsilon}_2 \stackrel{\sim}{\phi} + \epsilon_2 N \frac{\partial \mathcal{H}_0 }{\partial \stackrel{\sim}{\chi}^{\top}}.
\end{eqnarray} 
Finally, the smeared Hamiltonian constraint gives 
\begin{eqnarray} 
\{ \epsilon \mathcal{H}_0,    \pi^{ij} \dot{h}_{ij} + \stackrel{\sim}{\chi}^{\top}  \dot{\stackrel{\sim}{\phi} }    - \dot{\stackrel{\sim}{\chi}}^{\top} \stackrel{\sim}{\phi}  -N \mathcal{H}_0 \} = \nonumber \\
\{ \epsilon \mathcal{H}_0,  \pi^{ij} \dot{h}_{ij} \}   = \nonumber \\
\{  \epsilon \mathcal{H}_0, \pi^{ij} \}  \dot{h}_{ij} + \{  \epsilon \mathcal{H}_0, \dot{h}_{ij} \} \pi^{ij}.
\end{eqnarray}
One can evaluate this expression either by integrating over time and integrating by parts, or using the Anderson-Bergmann velocity Poisson bracket.  Either way, one finds the result\footnote{Previously I used a partial time derivative notation \cite{GRChangeNoKilling}, as did Anderson and Bergmann \cite{AndersonBergmann}.   But looking at what the Anderson-Bergmann Poisson bracket (alas, still poorly understood it seems) needs to mean to give correct answers rather than nonsense, a total derivative, recognizing both explicit time dependence through $\epsilon$ and implicit time dependence through phase space variables, is needed. }  
\begin{eqnarray} \{ \epsilon(t) \mathcal{H}_0, \pi^{ij} \dot{h}_{ij} - N \mathcal{H}_0 \} = \epsilon \dot{\mathcal{H}}_0 - \frac{ d}{ dt} \left(\epsilon \pi^{ij} \frac{ \partial \mathcal{H}_0 }{ \partial \pi^{ij} } \right). 
\end{eqnarray}

Plausibly the constraints inherited from vacuum GR, namely $\pi_N$ and $\mathcal{H}_0,$ contribute to the gauge generator $G$ in the same way as in GR:  
\begin{eqnarray} 
G[\epsilon,\dot{\epsilon}] = \dot{\epsilon} \pi_N + \epsilon \mathcal{H}_0+ \ldots, 
\end{eqnarray}
where the ellipsis stands for possible spinor terms.  Checking how just the old terms from GR change the canonical action can provide clues about possible new spinor terms.  One has
\begin{eqnarray} 
\{ \dot{\epsilon} \pi_N + \epsilon \mathcal{H}_0, \mathcal{L}_c \} = 
 \frac{ d}{ dt} \left(  \epsilon \mathcal{H}_0 -   \epsilon \pi^{ij} \frac{ \partial \mathcal{H}_0 }{ \partial \pi^{ij} } \right)   - \epsilon \dot{\stackrel{\sim}{\chi} }^{\top} \frac{\partial \mathcal{H}_0 }{\partial \stackrel{\sim}{\chi} }    -  \epsilon \frac{\partial \mathcal{H}_0 }{\partial \stackrel{\sim}{\phi} } \dot{\stackrel{\sim}{\phi} }.  
\end{eqnarray}
Thus there are some terms apt for cancellation by mixing in the primary (second-class) constraints from the spinors. We have already seen that $\mathcal{H}_0$ is not first-class and required contributions from the spinor constraints to give a first-class constraint $\bar{\mathcal{H}}_0$.

Adding in the spinor primary constraints with arbitrary phase space-independent coefficients gives
\begin{eqnarray} 
\{  \dot{\epsilon} \pi_N + \epsilon \mathcal{H}_0 +  (\pi_{\phi} -\stackrel{\sim}{\chi}^{\top} ) \epsilon_1 + \epsilon_2 (\pi_{\chi} + \stackrel{\sim}{\phi}), \mathcal{L}_c \} = \nonumber  \\
  \frac{ d}{ dt} \left(  \epsilon \mathcal{H}_0 -   \epsilon \pi^{ij} \frac{ \partial \mathcal{H}_0 }{ \partial \pi^{ij} } \right)
  - \epsilon \dot{\stackrel{\sim}{\chi}}^{\top} \frac{\partial \mathcal{H}_0 }{\partial \stackrel{\sim}{\chi} }      -   \epsilon \dot{\stackrel{\sim}{\phi}}^{\top} \frac{\partial \mathcal{H}_0 }{\partial \stackrel{\sim}{\phi} }  -
\dot{\epsilon}_1 \stackrel{\sim}{\chi}^{\top}  + \epsilon_1 \dot{\stackrel{\sim}{\chi}}^{\top} + N \epsilon_1 \frac{\partial \mathcal{H}_0 }{\partial \stackrel{\sim}{\phi}} - \epsilon_2 \dot{\stackrel{\sim}{\phi}} + \dot{\epsilon}_2  \stackrel{\sim}{\phi}  + N \epsilon_2 \frac{\partial \mathcal{H}_0 }{\partial \stackrel{\sim}{\chi}^{\top} }.
\end{eqnarray}
Cancelling the unwanted terms requires setting $\epsilon_1 = \frac{ \epsilon \dot{\stackrel{\sim}{\phi}}}{N}$ and $\epsilon_2 = \frac{ \epsilon \dot{\stackrel{\sim}{\chi}}}{N}.$
Thus one arrives at a change in the canonical Lagrangian that is just a total derivative, thus preserving Hamilton's equations: 
\begin{eqnarray} 
\frac{ d}{ dt} \left(  \epsilon \mathcal{H}_0 -   \epsilon \pi^{ij} \frac{ \partial \mathcal{H}_0 }{ \partial \pi^{ij} } \right) + 
\frac{d}{dt} \left[\frac{\epsilon}{N} (\dot{\stackrel{\sim}{\chi}}^{\top} \stackrel{\sim}{\phi} - \stackrel{\sim}{\chi} \dot{\stackrel{\sim}{\phi}}) \right]_.
\end{eqnarray} 
This expression apparently results from the Poisson bracket with the following expression as the gauge generator $G$:  
\begin{eqnarray} 
G= \dot{\epsilon} \pi_N + \epsilon \mathcal{H}_0 + (\pi_{\phi} - \stackrel{\sim}{\chi}^{\top} ) \frac{ \epsilon \dot{\stackrel{\sim}{\phi} }}{N} + \frac{ \epsilon \dot{\stackrel{\sim}{\chi}}}{N} (\pi_{\chi} - \stackrel{\sim}{\phi}) =\nonumber \\
\dot{\epsilon} \pi_N + \epsilon [ \mathcal{H}_0 + (\pi_{\phi} - \stackrel{\sim}{\chi}^{\top} ) \frac{  \dot{\stackrel{\sim}{\phi} }}{N} + \frac{  \dot{\stackrel{\sim}{\chi}}}{N} (\pi_{\chi} - \stackrel{\sim}{\phi})].
\end{eqnarray}

Clearly it is peculiar that the gauge generator contains velocities of the canonical variables.  This  peculiarity is shared by the original Anderson-Bergmann gauge generator for GR \cite{AndersonBergmann} and closely related to a matter that Bergmann found deeply problematic until he took Dirac to have resolved it  \cite{BergmannKomar,BergmannFadingWorldPoint,BergmannEarly,SalisburySyracuse1949to1962,BergmannObservables}.  But spinors, especially due to being linear rather than quadratic in derivatives in their field equations, have many peculiarities, so this one might be tolerable.  One would want to reflect on how this result bears on the quest for gauge transformations that are projectable to phase space in the sense of lacking velocities of the lapse, shift vector, electrostatic scalar potential and its Yang-Mills analogs, \emph{etc.} \cite{PonsSalisburyShepleyYang}. The alternative gauge generator $\bar{G}$ found below is more suited to phase space.

There is a gap, however, in this derivation:  initially the smearing functions were all assumed to be independent of the phase space quantities, but ultimately the smearing functions were made to depend on phase space variables after all.  So one must revisit the derivation using the final forms  $\epsilon_1 = \frac{ \epsilon \dot{\stackrel{\sim}{\phi}}}{N}$ and $\epsilon_2 = \frac{ \epsilon \dot{\stackrel{\sim}{\chi}}}{N},$   which have some nonzero Poisson brackets.  Fortunately the only nonzero brackets are with the canonical momenta associated with the lapse and the spinors, canonical momenta that in fact are absent from the canonical action.  Hence the extra terms vanish after all.  Thus the candidate for the gauge generator $G$ is vindicated:  its Poisson bracket with the canonical Lagrangian really does change the canonical Lagrangian by only a total derivative, thus preserving the Euler-Lagrange equations, that is, Hamilton's equations.  The arbitrary (nonvanishing) function $\epsilon(t)$ describes infinitesimal coordinate transformations for solutions of the spatially homogenized Einstein-Dirac equations.


\section{Role of Second-Class Constraints in Gauge Generator $G$?}

This example might be a useful one to examine in the context of the 1990s controversy about what role second-class constraints might play in the gauge generator.  One strategy, suggested by Castellani in the 1980s, is simply to eliminate second-class constraints using Dirac brackets.  
\begin{quote} 
Notice that the whole game is insensitive to the presence of second-class
constraints. Their effect is really to reduce the phase-space of the theory, not to
generate any symmetry. In any case, they can be eliminated by using Dirac brackets
[1].  \cite{CastellaniGaugeGenerator}   
\end{quote}

Supposing that we do not employ  Castellani's proposal for the moment,  the question returns:  what role, if any, do second-class constraints play in the gauge generator $G$ in ``mixed'' systems, those with both first-class and second-class constraints? 
 Sugano and Kimura
\begin{quote} 
\ldots   propose a method to construct the
generator $G$ of the infinitesimal gauge transformation
leaving the action   \vspace{-.25in} $$\hspace{1.5in} S= \int dt L (q, \dot{q}, t) \hspace{.6in}  (3.1)  \vspace{-.15in}$$
quasi-invariant. $G$ for special Lagrangians was first obtained
by Anderson and Bergmann. For the dynamical
system having FCC's alone, $G$ can be given by a linear
combination of the FCC's (Refs. 3, 4, 8, and 9). For the
system containing FCC's and SCC's, we can also analogously
construct $G$, though the method is rather complicated.
In such a case, $G$ turns out in general to be a
linear combination of the FCC's and SCC's. \cite{SuganoKimuraFirstSecond} 
 \end{quote} 
Our experience with the spinor example in this paper seems to align nicely with this description.  In particular instances some simplification is possible. ``If the first-class constraints and the Hamiltonian are in involution, the generator of pure gauge transformations can be obtained using only the first-class constraints.''  \cite{SuganoKimuraFirstSecond}
 
But Chitaia \emph{et al.} deny that second-class constraints ever play a role. 
\begin{quote}
In papers [4-9] it is queried, and in Refs. [17-19] it is even
asserted that second-class constraints also contribute to a
generator of gauge transformations which become global in
the absence of first-class constraints [17]. The generalized
Hamiltonian dynamics of systems with constraints of first
and second class has been studied relatively weakly up to
now.  \cite{ChitaiaGaugeFirstSecond} 
\end{quote}  
One should be able to decide the matter by calculation. 
\begin{quote}  To elucidate the role of second-class
constraints in local-symmetry transformations, we consider
first- and second-class constraints on the same basis in the hypothetical generator of these transformations. We prove
that the second-class constraints do not contribute to the
local-symmetry transformation law and, thus, the transformation
generator is a linear combination of only the first-class
constraints. \cite{ChitaiaGaugeFirstSecond} 
\end{quote} 
Commendably they employ  physically interesting examples including Chern-Simons theory and  spinor electrodynamics.  
Having noted the relevance of the spatially homogenized Einstein-Dirac system for this debate, I leave further study of it for another time.

What if we try to take Castellani's advice and use the Dirac brackets to eliminate the second-class constraints?  Let us conjecture that the resulting gauge generator $\bar{G}$ will be built out of the first-class constraints from the redefined set in the same way that the gauge generator for vacuum GR:  \begin{eqnarray}  \bar{G} = \epsilon \bar{\mathcal{H}}_0 +  \dot{\epsilon} \pi_N. \end{eqnarray} 
With the second-class constraints taken as identities, the new terms in $\bar{\mathcal{H}}$ disappear, while $\stackrel{\sim}{\chi}^{\top}$ is now canonically conjugate to $\stackrel{\sim}{\phi}$ up to a factor of $2.$ Let the resulting expression for the Hamiltonian constraint be called $\mathcal{H}_0|,$ where the $|$ indicates that one identifies $\stackrel{\sim}{\chi}^{\top}$ and $\pi_{\phi}.$  
 Let us ascertain what this new candidate gauge generator $\bar{G}$ does to the canonical Lagrangian:
\begin{eqnarray} 
 \{ \bar{G}, \mathcal{L}_c \}^* = \frac{d}{dt} \left( \epsilon \mathcal{H}_0| - \epsilon \pi^{ij} \frac{\partial \mathcal{H}_0| }{\partial \pi^{ij} }    
 - \frac{\epsilon}{2}  \stackrel{\sim}{\phi} \frac{\partial \mathcal{H}_0| }{\partial \stackrel{\sim}{\phi} }  - \frac{\epsilon}{2} \pi_{\phi} \frac{\partial \mathcal{H}_0| }{\partial \pi_{\phi}  } \right)_,
\end{eqnarray} 
where use has been made of the Anderson-Bergmann velocity Poisson bracket and also of the Dirac bracket  $\{ \stackrel{\sim}{\phi}, \pi_{\phi} \}^* = \frac{1}{2} I_{4\times 4}$.  As usual, the key point is that the resulting expression is a total derivative.  
Thus Castellani's briefly suggested procedure has worked:  eliminating the second-class constraints using Dirac brackets has yielded a gauge generator $\bar{G}$ composed entirely of first-class constraints and taking the same form as in vacuum GR.  If one wants gauge transformations to be projectable to phase space, that is, to lack velocities that cannot be replaced by canonical momenta, then one is further pleased that this gauge generator $\bar{G}$ lacks velocities:  the velocities were in the coefficients of the second-class primary spinor constraints, terms which disappeared when those constraints were taken as identities.  Thus at least in this case one can make changes of time coordinate much more phase space-friendly by eliminating the second-class constraints.  


\section{Change in  Einstein-Dirac Theory} 

In the literature on exact solutions of Einstein's equations \cite[p. 275]{ExactSolns2} one calls a solution ``stationary'' if (and where-when)  has a time-like Killing vector field.
A Killing vector field is a vector field such that the Lie derivative of the metric vanishes.  In that case one can adapt coordinates to that vector field so that the metric is independent of the coordinate $x$ for which the Killing vector field $\xi^{\mu}$ is $\frac{\partial}{\partial x}$ \cite[p. 99]{ExactSolns2}.  If the Killing vector field is  time-like, then it is plausible to use that coordinate as time.  Hence stationarity means that there exists a time coordinate such that the metric is independent of it.  While one can introduce apparent  time-dependence by using some other time coordinate that  wiggles in comparison to this one, the time-like Killing vector field shows that such apparent time dependence is phony.  Hence ``stationary'' and ``unchanging'' are synonyms for vacuum GR, when the metric is the only thing that exists and hence the only possible locus of change.  Thus an earlier paper spoke of change from the lack of a time-like Killing vector field, deploying in a Hamiltonian context standard conceptions from the exact solutions literature \cite{GRChangeNoKilling}. 
For a spatially homogeneous theory, one still has the freedom to bunch up or spread out the slices relative to physical (proper) time: reparametrization invariance.   That paper went on to consider the introduction of a real scalar field for matter, an example that, having no additional gauge freedom, adds no conceptual complications in the way that, say, Maxwell's electromagnetism does.

As discussed above, whereas the typical orthonormal tetrad formalism for the Einstein-Dirac equation introduces additional gauge freedom and thus adds conceptual complications regarding symmetries \cite{OrtinLieLorentzDerivatives}, eliminating that gauge freedom using a nonlinear group realization formalism removes those complications.  That is true whether one uses the symmetric tetrad condition developed by Ogievetsky and Polubarinov consciously into an alternative formalism, or one uses the Schwinger time gauge to fix the local boosts and the Kibble-Deser symmetric triad condition to fix the local rotations.

 In the case of spatial homogeneity considered in this paper, in which the shift vector plays no role and hence may be taken to vanish, those two nonlinear group realizations coincide.  Consequently one may borrow results from one to the other.  The Lie derivative of a spinor has been defined using the symmetric tetrad \cite{OPspinor}.  It involves a new term that vanishes when the metric tensor commutes (as a matrix) with the symmetric part of the gradient of the vector field describing the coordinate transformation.  For the example at hand, only the time component of that vector field is nonzero and it depends only on time, making the gradient of the descriptor vector field have only a $00$ component.  The vanishing shift vector implies that the metric has no time-space components, so only the time-time component contributes.  Consequently the metric and the symmetrized gradient of the transformation \emph{do} commute, annihilating the extra term.  Thus in our example the Lie derivative of the spinor is just the Lie derivative of a scalar.  That result harmonizes with the gauge generator $G$'s acting on the spinor fields just as it would on a scalar, as shown above.

How, if at all, can one generalize the vacuum GR definition of change as the lack of a time-like Killing vector field?  In the Einstein-Dirac system, real change, that is, ineliminable time dependence, could appear in the metric, in the spinor, in both, or just maybe in the relation between the two (supposing that one could eliminate change from the metric in one coordinate system and from the spinor in a different coordinate system, but there was no coordinate system eliminating change from both the metric and the spinor).  It turns out to be easier to define the absence of change, and then to define change as its negation.  
For there to be no change, it suffices that there exist a vector field such that the Lie derivative of the pair $\langle g_{\mu\nu}, \psi \rangle$ vanishes.  This pair very nearly comprises a nonlinear geometric object \cite{GRChangeNoKilling} in the classical sense \cite{SzybiakLie,SzybiakCovariant,Tashiro1,Tashiro2} of a set of components and a transformation law.  For nonlinear geometric objects, the Lie derivative is not itself a geometric object in general, but the Lie derivative along with the object itself \emph{is} a geometric object.  The pair  $\langle g_{\mu\nu}, \psi \rangle$ is not quite a geometric object due to spinor double-valuedness and due to the inadmissibility of coordinate systems that give negative eigenvalues to the matrix $g_{\mu\nu} \eta^{\nu\rho}$ (where $\eta^{\nu\rho}=diag(-1,1,1,1)$) \cite{BilyalovConservation,BilyalovSpinors,PittsSpinor,TAM2013TimeandFermionsConfProc},  
 but neither limitation is relevant to transformations near the identity, which the Lie derivative uses.  Hence  $\langle g_{\mu\nu}, \psi \rangle$ is close enough to a nonlinear geometric object, and $\langle g_{\mu\nu}, \psi, \pounds_{\xi} g_{\mu\nu}, \pounds_{\xi} \psi \rangle$ is close enough to a nonlinear geometric object as well.  It is unusual, however, in that it has a part that is itself a geometric object, indeed a linear geometric object, namely the metric (or some equivalent entity or, if one strips away unnecessary elements, the conformal part of the metric); the Lie derivative of the metric is also a linear geometric by itself.  Importantly for present purposes, the transformation rule for this $4$-piece spinorial almost-geometric object is strictly linear in the spinor field, making $0$ spinor an invariant notion.

One can in fact define change for GR with any matter fields that have no gauge `group' besides spatio-temporal coordinate transformations.  For convenience I  restrict matter fields to those having coordinate transformations depending only on the first derivatives of the coordinates:  $\frac{\partial x^{\mu^{\prime}}  }{\partial x^{\alpha}}$ but not $x^{\mu}$ or  $\frac{\partial^2 x^{\mu^{\prime} }}{\partial x^{\alpha} \partial x^{\beta} }$. This assumption includes almost every physically interesting theory; even many theories with an independent connection (hence with second derivatives in the transformation law) can be reformulated using the difference between that independent connection and the Levi-Civita connection, which is a tensor.  Hence though the assumption could be lifted in Tashiro's formalism, it is not very restrictive and leads to considerable simplification.  I hope to include internal gauge groups on another occasion; but this paper shows that spinors can be treated already due to the nonlinear formalism.

Following Tashiro \cite{Tashiro1}, one can define the Lie derivative for the complex $\Omega^{\mathcal{A}},$ which I will specialize as $\langle g_{\alpha\beta}, \psi \rangle$ on occasion. (Whereas spinors were treated as column  matrices ($\stackrel{\sim}{\phi}$ and $\stackrel{\sim}{\chi}$) or row matrices above, now they are subsumed into index notation for the purpose of differential geometry.)  This argument works for any collection of tensors, tensor densities, or whatever else has a transformation rule involving at most the first derivative of the coordinates (and not involving the coordinates themselves) that has no additional gauge freedom---which even includes spinors if  one uses a nonlinear group realization formalism.  (In fact the restriction to avoiding second derivatives  in the transformation rule is not necessary, but it covers most interesting cases and streamlines the exposition.)   $\Omega^{\mathcal{A}} $ is a (potentially) nonlinear geometric object, or at least behaves like one for coordinate transformations that are not too large (such as swapping $t$ and $x$ or rotating by $ 2 \pi$, for example).  
 Gathering all the fields into an ordered $n$-tuple $\Omega^\mathcal{A},$ under a coordinate transformation one has 
\begin{eqnarray} 
\Omega^{\mathcal{A}\prime} = F^\mathcal{A}\left(\Omega^{\mathcal{M}}, \frac{\partial x^{\mu\prime} }{\partial x^{\nu} }\right) = \langle g_{\mu\nu} \frac{\partial x^{\mu}}{\partial x^{\alpha\prime} } \frac{\partial x^{\nu}}{\partial x^{\beta\prime} } , F^A_M \left( \frac{\partial x^{\mu\prime}}{\partial x^{\alpha} }, g_{\alpha\beta} \right) \psi^M \rangle. 
\end{eqnarray}
$F^A_M$ of course includes $\gamma^A$ matrices, but these, being numerical, do not need to be listed explicitly.  
The Lie derivative is defined involving a term differentiating $ F^{\mathcal{A}}$ with respect to $ \frac{\partial x^{\mu\prime}}{\partial x^{\alpha}}$ :  
\begin{eqnarray} F^{\mathcal{A}\nu}_{\lambda} = \left. \frac{ \partial F^{\mathcal{A} }}{\partial \frac{\partial x^{\lambda\prime}}{\partial x^{\nu} }}   \right| \frac{\partial x^{\lambda\prime}}{\partial x^{\nu} } \rightarrow \delta^{\lambda}_{\nu}.
\end{eqnarray}
The Lie derivative is then \begin{eqnarray} \pounds_{\xi} \Omega^{\mathcal{A}} = \xi^{\lambda} \Omega^{\mathcal{A}},_{\lambda} - \xi^{\lambda},_{\nu} F^{\mathcal{A}\nu}_{\lambda}(\Omega).
\end{eqnarray} 
It is straightforward to show that this formula, a special case of Tashiro's, gives the usual formulas for vectors, covectors, tensors, \emph{etc.}  The Lie derivative has the coordinate transformation rule 
\begin{eqnarray} 
\pounds_{\xi} \Omega^{\mathcal{A}} = \frac{ \partial F^{\mathcal{A}\prime}}{\partial \Omega^{\mathcal{B}}} \pounds_{\xi} \Omega^{\mathcal{B}}.
\end{eqnarray}
For a strictly linear or affine (linear inhomogeneous) geometric object, the Lie derivative itself is a geometric object, one of the same type as $\Omega^{\mathcal{A}}$  or in some cases a bit nicer (such as strictly linear instead of affine).  
Applying this technology to  $\Omega^{\mathcal{A}} = \langle g_{\alpha\beta}, \psi \rangle$, one finds that the Lie derivative includes two pieces, the Lie derivative of the metric (which is a rank $2$ covariant tensor as usual) and the Lie derivative of the spinor, which is not a geometric object; it does not even have the same transformation properties as the spinor $\psi.$  The transformation rule of the Lie derivative of the spinor field involves both that Lie derivative itself \emph{and} the Lie derivative of the metric (or, if one is more frugal and judicious in choosing density weights, the Lie derivative of the conformal part of the metric \cite{PittsSpinor,TAM2013TimeandFermionsConfProc}, making conformal/Weyl invariance manifest by omitting any volume element).  This fact explains the common belief that the Lie derivative of a spinor only makes sense for Killing vectors \cite{BennTucker} or for conformal Killing vectors \cite[p. 101]{PenroseRindler2}:  the transformed Lie derivative of the spinor is only proportional to the original Lie derivative of the spinor when the Lie derivative of the (conformal part of the) metric vanishes.

One is now in a position to define change for physical theories including any number of fields with coordinate transformation rules involving only first derivatives (of the coordinates with respect to other coordinates) and no additional gauge freedom.  There is \emph{no change} just in case there exists a time-like\footnote{If a theory either lacks the resources to call a vector field time-like or yields an ambiguous verdict due to a multiplicity of such resources (such as having two metrics that are not conformally related) that are considered equally relevant, then change is either meaningless or ambiguous, respectively; that is no flaw in the definition.}  vector field $\xi^{\mu}$ such that  $\pounds_{\xi} \Omega^{\mathcal{A}}=0.$  Assuming that $\Omega^{\mathcal{A}}$ contains a space-time metric, this condition is logically stronger than being a Killing vector field, because the field needs to be `Killing' (so to speak) for all the other fields as well.  Given that $\langle \Omega^{\mathcal{A}},  \pounds_{\xi} \Omega^{\mathcal{A}} \rangle$ has the transformation rule shown above,  $\pounds_{\xi} \Omega^{\mathcal{A}}=0$   is a coordinate-invariant condition.  Given that this condition holds in any coordinate system, it holds in particular in a coordinate system in which in a neighborhood $\xi^{\mu} = (1,0,0,0)$.  Such a coordinate system exists locally for \emph{any} (sufficiently smooth\ldots) vector field \cite{BergmannConservation}; if the coordinate is time-like relative to a metric, then it is reasonable to call that coordinate ``time.''  In such a coordinate system   $ \pounds_{\xi} \Omega^{\mathcal{A}} = \dot{\Omega}^{\mathcal{A}}=0$.  Hence the existence of this stronger-than-Killing vector field $\xi^{\mu}$ implies that there exists a coordinate system in which  $\Omega^{\mathcal{A}}$ is independent of the coordinate adapted to  $\xi^{\mu}$ and \emph{vice versa}. The existence of such a vector field that is also time-like is thus the criterion for the absence of change.  Consequently change is the non-existence of a time-like stronger-than-Killing vector field such that  $ \pounds_{\xi} \Omega^{\mathcal{A}} = 0$.
This definition works for quite a variety of physical theories, albeit excluding Maxwell's electromagnetism and Yang-Mills due to their internal gauge groups. For electromagnetism, it might be satisfactory to use $F_{\mu\nu}$ and avoid addressing the gauge freedom, but this option would not work with local fields for Yang-Mills.   Using nonlinear group realizations, however, spinor fields \emph{are} included in the collection of theories with no additional internal gauge symmetry.  Thus one can understands change  in the Einstein-Dirac system already.


\section{Change from $G$ in Spatially Homogeneous Einstein-Dirac}

Dropping all dependence on space once more, one can calculate what the gauge generator $G$ generates when acting on the various quantities in phase space. One finds the following, feeling free to use Hamiltonian equations of motion as needed (``on shell'') in some cases: 
\begin{eqnarray} 
\{ G, \stackrel{\sim}{\phi} \} = - \frac{\epsilon \dot{ \stackrel{\sim}{\phi} } }{N}, \nonumber \\
\{G, \stackrel{\sim}{\chi}^{\top} \} = - \frac{\epsilon \dot{ \stackrel{\sim}{\chi} } }{N}, \nonumber \\
 \{ G,  \pi_{\phi} \} = \epsilon \frac{\partial \mathcal{H}_0 }{\partial \stackrel{\sim}{\phi} } + \frac{ \epsilon \dot{ \stackrel{\sim}{\chi}}^{\top} }{N} =    - \frac{\epsilon \dot{ \pi}_{\phi}  }{N}      \hspace{.15in} (on \hspace{.08in} shell), \nonumber \\
 \{ G,  \pi_{\chi} \} = \epsilon \frac{\partial \mathcal{H}_0 }{\partial \stackrel{\sim}{\phi} } - \frac{ \epsilon \dot{ \stackrel{\sim}{\phi} }}{N} =    - \frac{\epsilon \dot{ \pi}_{\chi}  }{N}      \hspace{.15in} (on \hspace{.08in} shell), \nonumber \\
\{ G, h_{ij} \} = \frac{\epsilon}{N} \{ \mathcal{H}_p, h_{ij} \}  = -\frac{\epsilon}{N} \dot{h}_{ij}   \hspace{.15in} (on \hspace{.08in} shell), \nonumber \\
\{ G, \pi^{ij} \} =  \frac{\epsilon}{N} \{  \mathcal{H}_p, \pi^{\ij} \} = -\frac{\epsilon}{N} \dot{\pi}^{ij}  \hspace{.15in} (on \hspace{.08in} shell), \nonumber \\
\{G, N \} = \{ \dot{\epsilon} \pi_N, N \} = -\dot{\epsilon}, \nonumber \\
\{ G, \pi_N \} = - (\pi_{\phi} - \stackrel{\sim}{\chi}^{\top} ) \frac{\epsilon}{N^2} \dot{\stackrel{\sim}{\phi} } - \dot{\stackrel{\sim}{\chi} } \frac{\epsilon}{N^2} (\pi_{\chi}+ \stackrel{\sim}{\phi}) = 0      \hspace{.15in} (on \hspace{.08in} shell).
\end{eqnarray} 
These results look promising in terms of equivalence (for solutions) to the $4$-dimensional Lie derivative, which arises from infinitesimal coordinate transformations.  One can calculate that $$ \delta ^{\mu}_0 \delta^{\nu}_ 0  \pounds_{\xi} g_{\mu\nu} = - 2N \epsilon,_0,$$ showing agreement when one recalls that $g_{00}= -N^2$ (in the absence of the shift vector) and that $\epsilon = - n_{\mu} \xi^{\mu} = N \xi^0.$   
One can also see that $$ \delta ^{\mu}_i \delta^{\nu}_ j  \pounds_{\xi} g_{\mu\nu} =   \dot{h}_{ij}.$$  Hence $G$ changes the space-time metric in accord with the Lie derivative for solutions.  
Above it was found that the Ogievetsky-Polubarinov Lie derivative of a spinor is exactly the Lie derivative required for the spatially homogeneous toy theory and that the Lie derivative of the spinor takes the same form as the Lie derivative of a scalar. As we have just seen, $G$ also generates a change of the spinors in the same form as on scalar.  (One should not confuse the \emph{spatial}  density weight $\frac{1}{2}$ of the spinors with their transformation properties under change of time coordinate, a largely separate one-dimensional tensor calculus.)  
Hence the gauge generator $G$ generates changes of the phase space variables that are equivalent to temporal coordinate transformations for solutions in the toy theory.


\section{Observables in Einstein-Dirac Theory}

If one imposes the condition of vanishing Poisson bracket with $G$, that is a condition not of appropriate behavior under a gauge transformation or of being an observable, but of being stationary, changeless, having a time-like stronger-than-Killing vector field.  Hence the reason that change has been missing in Hamiltonian GR is that a condition of changelessness has been imposed and mistaken for a condition of appropriate behavior for an observable.  But observables should be only covariant, not invariant, under coordinate transformations, in the sense of changing by a Lie derivative that has a group property, not of having vanishing Lie derivative \cite{GRChangeNoKilling,ObservablesEquivalentCQG}.  The condition of vanishing Lie derivative (or \emph{a forterioi} vanishing Poisson bracket with all $8$ first-class constraints involved in the spatial and temporal gauge generators \cite{CastellaniGaugeGenerator}) requires that observables be the same at different space-time points with the same coordinate value in different coordinate systems, akin to 1a.m. standard time \emph{vs.} 1 a.m. daylight savings time, a condition that obviously implies changelessness and has nothing to do with observability. Given some of the roots of Bergmann's concept of observables, one should not expect a close connection between them and observables in the ordinary sense of phenomena that can be observed, Kiefer has noted \cite[pp. 105, 143]{Kiefer3rd}.  

Using the principle that equivalent theories (or theory formulations) should have equivalent observables, I argued that observables in GR should be invariant under internal gauge symmetries (which is not novel) but only covariant under coordinate transformations, so observables are basically geometric objects, or tensor calculus all over again, such as $g_{\mu\nu}$ and its concomitants \cite{ObservablesEquivalentCQG,ObservablesLSEFoP,ObservablesEinsteinMaxwellFoP}.  The nonlinear group realization formalism comes very close to including spinors as part of a geometric object along with the metric (or its conformal part). Does it follow the  spinors are observables or rather that $\langle g_{\mu\nu}, \psi \rangle$ is an observable?  That cannot be true for the usual reason from particle physics:  spinors change sign under a rotation by $2 \pi.$ Nothing observable (in the ordinary sense) would do that.  Hence the particle physics qualification on observing spinors will of course apply in Hamiltonian GR:  observables should instead be suitable bilinear expressions in the spinor(s), expressions which are unchanged under rotation by $2\pi.$  Ogievetsky and Polubarinov exhibit various bilinear expressions in the symmetric tetrad nonlinear group realization formalism \cite{OPspinor}.  In this present paper's toy theory with vanishing shift and no spatial dependence, that formalism is the same as the one using the Schwinger time gauge and the Kibble-Deser symmetric triad, conditions that one might reasonably regard as the Arnowitt-Deser-Misner split of the symmetric square root of the metric.


\section{Conclusion}

Using a nonlinear realization of the `group' (roughly speaking) of space-time coordinate transformations, one can express the Einstein-Dirac equation with spinors almost fitting into the realm of classical geometrical objects, thus having a classical Lie derivative along any vector field with no extra gauge group.  Such a formulation is here simplified by the imposition of the simplest form of spatial homogeneity in order to focus attention on changes of time coordinate, a matter often neglected in Hamiltonian formalisms.  A gauge generator (or two) for change of time coordinate that changes the canonical Lagrangian by at most a total derivative, thus preserving the Euler-Lagrange (Hamiltonian) equations was found.  Change was found to be the absence of a stronger-than-Killing vector for which the Lie derivative of the metric and of the spinor together vanishes.  For observables one expects spinor fields to require bilinearity to avoid a change of sign under $2 \pi$ rotations.  At least apart from quantization, change is not missing or paradoxical in GR even with (commuting) spinorial matter.

\section{Acknowledgments}

This work was supported by the National Science Foundation (USA) Science, Technology and Society grant \#1734402  with  Claus Kiefer and Jeremy Butterfield.  


\begin{thebibliography}{100}

\bibitem{AndersonChange}
James~L. Anderson.
\newblock Absolute change in general relativity.
\newblock In {\em Recent Developments in General Relativity}, pages 121--126.
  Pergamon and PWN, Oxford and Warsaw, 1962.

\bibitem{IshamTime}
C.~J. Isham.
\newblock Canonical quantum gravity and the problem of time.
\newblock In L.~A. Ibort and M.~A. Rodr\'{i}guez, editors, {\em Integrable
  Systems, Quantum Groups, and Quantum Field Theories}, pages 157--287. Kluwer,
  Dordrecht, 1993.
\newblock {Lectures presented at the NATO Advanced Study Institute ``Recent
  Problems in Mathematical Physics,'' Salamanca, June 15-27, 1992;
  gr-qc/9210011}.

\bibitem{BelotEarman}
Gordon Belot and John Earman.
\newblock {Pre-Socratic} quantum gravity.
\newblock In Craig Callender and Nick Huggett, editors, {\em Philosophy Meets
  Physics at the Planck Scale}, pages 213--255. Cambridge University Press,
  Cambridge, 2001.

\bibitem{EarmanMcTaggart}
John Earman.
\newblock Thoroughly modern {McTaggart: Or, What McTaggart} would have said if
  he had read the {General Theory of Relativity}.
\newblock {\em Philosophers' Imprint}, 2(3), 2002.
\newblock http://www.philosophersimprint.org/.

\bibitem{RicklesTimeStructureQG}
Dean Rickles.
\newblock Time and structure in canonical gravity.
\newblock In Dean Rickles, Steven French, and Juha Saatsi, editors, {\em
  Structural Foundations of Quantum Gravity}, pages 152--195. Clarendon Press,
  Oxford, 2006.

\bibitem{HuggettWuthrichTimeQG}
Nick Huggett, Tiziana Vistarini, and Christian W\"{u}thrich.
\newblock Time in quantum gravity.
\newblock In Adrian Bardon and Heather Dyke, editors, {\em The Blackwell
  Companion to the Philosophy of Time}, pages 242-- 261. Blackwell, Chichester,
  2013.

\bibitem{AndersonRoyaumont}
James~L. Anderson.
\newblock Generation of coordinate conditions and the construction of
  invariants in covariant theories.
\newblock In {\em Les Th\'{e}ories Relativistes de la Gravitation, Royaumont,
  21-27 Juin 1959}, pages 373--384. Centre National de la Recherche
  Scientifique, 1962.

\bibitem{KucharCanonical93}
Karel~V. Kucha\v{r}.
\newblock Canonical quantum gravity.
\newblock In R.~J. Gleiser, C.~N. Kozameh, and O.~M. Moreschi, editors, {\em
  General Relativity and Gravitation 1992: Proceedings of the Thirteenth
  International Conference on General Relativity and Gravitation held at
  Cordoba, Argentina, {28 June--4 July} 1992}, pages 119--150. Institute of
  Physics Publishing, Bristol, 1993.
\newblock arXiv:gr-qc/9304012.

\bibitem{McTaggart}
John~Ellis McTaggart.
\newblock The unreality of time.
\newblock {\em Mind: A Quarterly Review of Psychology and Philosophy},
  17:457--474, 1908.

\bibitem{MaudlinMcTaggart}
Tim Maudlin.
\newblock Thoroughly muddled {McTaggart: Or, How} to abuse gauge freedom to
  generate metaphysical monstrosities.
\newblock {\em Philosophers' Imprint}, 2(4), 2002.
\newblock http://www.philosophersimprint.org/, with a reply by John Earman.

\bibitem{EarmanOde}
John Earman.
\newblock Getting a fix on gauge: An ode to the constrained {Hamiltonian}
  formalism.
\newblock In Katherine Brading and Elena Castellani, editors, {\em Symmetries
  in Physics: Philosophical Reflections}, pages 140--162. Cambridge University
  Press, Cambridge, 2003.

\bibitem{HealeyGRchangelessincoherent}
Richard Healey.
\newblock Can physics coherently deny the reality of time?
\newblock In Craig Callender, editor, {\em Time, Reality \& Experience}, pages
  293--316. Cambridge University Press, Cambridge, 2002.
\newblock Royal Institute of Philosophy Supplement 50.

\bibitem{PonsDirac}
Josep~M. Pons.
\newblock On {Dirac's} incomplete analysis of gauge transformations.
\newblock {\em Studies in History and Philosophy of Modern Physics},
  36:491--518, 2005.
\newblock arXiv:physics/0409076v2.

\bibitem{BarbourFosterPrimary}
Julian Barbour and Brendan~Z. Foster.
\newblock Constraints and gauge transformations: {Dirac's} theorem is not
  always valid.
\newblock 2008.
\newblock arXiv:08081223[gr-qc].

\bibitem{FirstClassNotGaugeEM}
J.~Brian Pitts.
\newblock A first class constraint generates not a gauge transformation, but a
  bad physical change: {The} case of electromagnetism.
\newblock {\em Annals of Physics}, 351:382--406, 2014.
\newblock arXiv:1310.2756.

\bibitem{GRChangeNoKilling}
J.~Brian Pitts.
\newblock Change in {Hamiltonian} general relativity from the lack of a
  time-like {Killing} vector field.
\newblock {\em Studies in History and Philosophy of Modern Physics}, 47:68--89,
  2014.
\newblock arXiv:1406.2665.

\bibitem{ObservablesEquivalentCQG}
J.~Brian Pitts.
\newblock Equivalent theories redefine {Hamiltonian} observables to exhibit
  change in {General Relativity}.
\newblock {\em Classical and Quantum Gravity}, 34(055008), 2017.
\newblock arXiv:1609.04812 [gr-qc].

\bibitem{Ohanian}
Hans Ohanian and Remo Ruffini.
\newblock {\em Gravitation and Spacetime}.
\newblock Norton, New York, second edition, 1994.

\bibitem{MukundaSymmetries}
N.~Mukunda.
\newblock Symmetries and constraints in generalized {Hamiltonian} dynamics.
\newblock {\em Annals of Physics}, 9:408--433, 1976.

\bibitem{MukundaGaugeGenerator}
N.~Mukunda.
\newblock Generators of symmetry transformations for constrained {Hamiltonian}
  systems.
\newblock {\em Physica Scripta}, 21:783--791, 1980.

\bibitem{SalisburySundermeyerEinstein}
Donald~C. Salisbury and Kurt Sundermeyer.
\newblock Realization in phase space of general coordinate transformations.
\newblock {\em Physical Review D}, 27:740--756, 1983.

\bibitem{PonsSalisburyShepley}
Josep Pons, Donald~C. Salisbury, and Lawrence~C. Shepley.
\newblock Gauge transformations in the {Lagrangian and Hamiltonian} formalisms
  of generally covariant theories.
\newblock {\em Physical Review D}, 55:658--668, 1997.
\newblock gr-qc/9612037.

\bibitem{ShepleyPonsSalisburyTurkish}
Lawrence~C. Shepley, Josep~M. Pons, and Donald~C. Salisbury.
\newblock Gauge transformations in general relativity---{A} report.
\newblock {\em Turkish Journal of Physics}, 24(3):445--452, 2000.
\newblock Regional Conference on Mathematical Physics IX, 9-14 August 1999,
  Istanbul, Turkey.

\bibitem{PonsSalisbury}
Josep~M. Pons and Donald~C. Salisbury.
\newblock Issue of time in generally covariant theories and the
  {Komar-Bergmann} approach to observables in general relativity.
\newblock {\em Physical Review D}, 71:124012, 2005.
\newblock arXiv:gr-qc/0503013.

\bibitem{PonsReduce}
Josep~M. Pons and Lawrence~C. Shepley.
\newblock Dimensional reduction and gauge group reduction in {Bianchi}-type
  cosmology.
\newblock {\em Physical Review D}, 58:024001, 1998.
\newblock arXiv:gr-qc/9805030.

\bibitem{SuganoExtended}
Reiji Sugano, Yusho Kagraoka, and Toshiei Kimura.
\newblock Gauge transformations and gauge-fixing conditions in constraint
  systems.
\newblock {\em International Journal of Modern Physics A}, 7:61--89, 1992.

\bibitem{SuganoGeneratorQM}
Reiji Sugano, Tohru Sohkawa, and Toshiei Kimura.
\newblock Relation between generators of gauge transformations and subsidiary
  conditions on state vectors. {Point} mechanical systems with arbitrary
  numbers of constraints.
\newblock {\em Progress of Theoretical Physics}, 73:1025--1042, 1985.

\bibitem{SuganoGaugeVelocityI}
Reiji Sugano and Yusho Kagraoka.
\newblock Extension to velocity dependent gauge transformations --- {I.
  General} form of the generator.
\newblock {\em Zeitschrift f\"{u}r Physik C}, 52:437--442, 1991.

\bibitem{SuganoGaugeGenerator}
Reiji Sugano, Yoshihiko Saito, and Toshiei Kimura.
\newblock Generator of gauge transformation in phase space and velocity phase
  space.
\newblock {\em Progress of Theoretical Physics}, 76:283--301, 1986.

\bibitem{SalisburySyracuse1949to1962}
Donald Salisbury.
\newblock Toward a quantum theory of gravity: {Syracuse} 1949-1962.
\newblock In Alexander Blum, Roberto Lalli, and J\"{u}rgen Renn, editors, {\em
  The Renaissance of General Relativity in Context}, volume~16 of {\em Einstein
  Studies}, pages 221--255. Birkh\"{a}user, 2020.
\newblock arXiv:1909.05412v1 [physics.hist-ph].

\bibitem{BergmannObservables}
J.~Brian Pitts.
\newblock {Peter Bergmann} on observables in {Hamiltonian} general relativity:
  {A} historical-critical investigation.
\newblock {\em Studies in History and Philosophy of Science}, Accepted subject
  to minor revisions, 2021.

\bibitem{DiracHamGR}
Paul A.~M. Dirac.
\newblock The theory of gravitation in {Hamiltonian} form.
\newblock {\em Proceedings of the Royal Society of London A}, 246:333--343,
  1958.

\bibitem{RosenfeldQG}
L.~Rosenfeld.
\newblock {Zur Quantelung der Wellenfelder}.
\newblock {\em Annalen der Physik}, 397:113--152, 1930.
\newblock Translation by {Donald Salisbury and Kurt Sundermeyer}, ``On the
  Quantization of Wave Fields,'' \emph{The European Physical Journal H} {\bf
  42} (2017), pp. 63-94.

\bibitem{AndersonBergmann}
James~L. Anderson and Peter~G. Bergmann.
\newblock Constraints in covariant field theories.
\newblock {\em Physical Review}, 83:1018--1025, 1951.

\bibitem{ObservablesLSEFoP}
J.~Brian Pitts.
\newblock Equivalent theories and changing {Hamiltonian} observables in
  {General Relativity}.
\newblock {\em Foundations of Physics}, 48:579--590, 2018.
\newblock doi:10.1007/s10701-018-0148-1; arXiv.org:1803.10059; PhilSci.

\bibitem{Nijenhuis}
Albert Nijenhuis.
\newblock {\em Theory of the Geometric Object}.
\newblock PhD thesis, University of Amsterdam, 1952.
\newblock Supervisor {Jan A. Schouten}.

\bibitem{Schouten}
Jan~A. Schouten.
\newblock {\em Ricci-Calculus: An Introduction to Tensor Analysis and Its
  Geometrical Applications}.
\newblock Springer, Berlin, second edition, 1954.

\bibitem{Anderson}
James~L. Anderson.
\newblock {\em Principles of Relativity Physics}.
\newblock Academic, New York, 1967.

\bibitem{ObservablesEinsteinMaxwellFoP}
J.~Brian Pitts.
\newblock What are observables in {Hamiltonian Einstein-Maxwell} theory?
\newblock {\em Foundations of Physics}, 49:786--796, 2019.
\newblock arXiv:1907.09473 [gr-qc].

\bibitem{WeylGravitationElectron}
Hermann Weyl.
\newblock Gravitation and the electron.
\newblock {\em Proceedings of the National Academy of Sciences of the United
  States of America}, 15:323--334, 1929.

\bibitem{WeylElektronGravitation}
Hermann Weyl.
\newblock Elektron und {Gravitation}.
\newblock {\em Zeitschrift f\"{u}r Physik}, 56:330--352, 1929.
\newblock Translation in {Lochlainn O'Raifeartaigh, \emph{The Dawning of Gauge
  Theory}, Princeton University Press, Princeton (1997), pp. 121-144.}

\bibitem{WeylRice}
Hermann Weyl.
\newblock Gravitation and the electron.
\newblock {\em The Rice Institute Pamphlet}, 16:280--295, 1929.

\bibitem{CartanSpinor}
{\'{E}lie}~Cartan and Andr\'{e} Mercier.
\newblock {\em The Theory of Spinors}.
\newblock Massachusetts Institute of Technology Press, Cambridge, 1966.
\newblock {French} original 1937.

\bibitem{OPspinor}
V.~I. Ogievetski\u{i} and I.~V. Polubarinov.
\newblock Spinors in gravitation theory.
\newblock {\em Soviet Physics JETP}, 21:1093--1100, 1965.

\bibitem{OP}
V.~I. Ogievetsky and I.~V. Polubarinov.
\newblock Interacting field of spin 2 and the {Einstein} equations.
\newblock {\em Annals of Physics}, 35:167--208, 1965.

\bibitem{PittsSpinor}
J.~Brian Pitts.
\newblock The nontriviality of trivial general covariance: {How}  electrons
  restrict `time' coordinates, spinors (almost) fit into tensor calculus, and
  $7/16$ of a tetrad is surplus structure.
\newblock {\em Studies in History and Philosophy of Modern Physics}, 43:1--24,
  2012.

\bibitem{BennTucker}
Ian~M. Benn and Robin~W. Tucker.
\newblock {\em An Introduction to Spinors and Geometry with Applications in
  Physics}.
\newblock Adam Hilger, Bristol, 1987.

\bibitem{PenroseRindler2}
Roger Penrose and Wolfgang Rindler.
\newblock {\em Spinors and Space-time, Volume 2: Spinor and Twistor Methods in
  Space-time Geometry}.
\newblock Cambridge University Press, Cambridge, 1986.

\bibitem{DeWittDissertation}
Carl~Bryce {Seligman [DeWitt]}.
\newblock {\em {I. The} Theory of Gravitational Interactions. {II. The}
  Interaction of Gravitation with Light}.
\newblock PhD thesis, Harvard University, 1949.
\newblock Supervisor {Julian Schwinger}.

\bibitem{DeWittSpinor1950}
Bryce~Seligman DeWitt.
\newblock On the application of quantum perturbation theory to gravitational
  interactions.
\newblock 1950.
\newblock https://repositories.lib.utexas.edu/handle/2152/9620.

\bibitem{DeWittSpinor}
Bryce~Seligman DeWitt and C\'{e}cile~Morette DeWitt.
\newblock The quantum theory of interacting gravitational and spinor fields.
\newblock {\em Physical Review}, 87:116--122, 1952.

\bibitem{DeWittDToGaF}
Bryce~S. DeWitt.
\newblock {\em Dynamical Theory of Groups and Fields}.
\newblock Gordon and Breach, New York, 1965.

\bibitem{BilyalovSpinors}
Ranat~F. Bilyalov.
\newblock Spinors on {Riemannian} manifolds.
\newblock {\em Russian Mathematics (Iz. VUZ)}, 46(11):6--23, 2002.

\bibitem{DeffayetSymmetricTetrad}
C.~Deffayet, J.~Mourad, and G.~Zahariade.
\newblock A note on ``symmetric'' vielbeins in bimetric, massive, perturbative
  and non perturbative gravities.
\newblock {\em Journal of High Energy Physics}, 1303(086), 2013.
\newblock arXiv:1208.4493 [gr-qc].

\bibitem{TenerifeProgressGravity}
J.~Brian Pitts.
\newblock Progress and gravity: Overcoming divisions between general relativity
  and particle physics and between science and hps.
\newblock In Khalil Chamcham, Joseph Silk, John Barrow, and Simon Saunders,
  editors, {\em The Philosophy of Cosmology}, pages 263--282. Cambridge
  University Press, Cambridge, 2017.
\newblock https://arxiv.org/abs/1907.11163.

\bibitem{KibbleSymmetric}
T.~W.~B. Kibble.
\newblock Canonical variables for the interacting gravitational and {Dirac}
  fields.
\newblock {\em Journal of Mathematical Physics}, 4:1433--1437, 1963.

\bibitem{DeserMoller}
Stanley Deser.
\newblock Note on {M}{\o}ller's gravitational stress tensor.
\newblock {\em Physics Letters}, 7:42--43, 1963.

\bibitem{DeserCargese}
Stanley Deser.
\newblock Dynamics of supergravity.
\newblock In Maurice L\'{e}vy and Stanley Deser, editors, {\em Recent
  Developments in Gravitation: Carg\`{e}se 1978}, pages 461--477. Plenum, New
  York, 1979.

\bibitem{MTW}
Charles Misner, Kip Thorne, and John~A. Wheeler.
\newblock {\em Gravitation}.
\newblock Freeman, New York, 1973.

\bibitem{GambiniPullin}
Rodolfo Gambini, Rafael Porto, Sebastian Torterolo, and Jorge Pullin.
\newblock Conditional probabilities with {Dirac} observables and the problem of
  time in quantum gravity.
\newblock {\em Physical Review D}, 79(041501), 2009.
\newblock arXiv:0809.4235 [gr-qc].

\bibitem{RovelliForgetTime}
Carlo Rovelli.
\newblock ``{Forget} time''.
\newblock {\em Foundations of Physics}, 41:1475--1490, 2011.

\bibitem{MukundaSamuelConstrainedGeometric}
G.~Marmo, N.~Mukunda, and J.~Samuel.
\newblock Dynamics and symmetry for constrained systems: {A} geometrical
  analysis.
\newblock {\em Rivista del Nuovo Cimento}, 6:1--62, 1983.

\bibitem{LusannaVelocityHamiltonian}
Luca Lusanna.
\newblock An enlarged phase space for finite-dimensional constrained systems,
  unifying their {Lagrangian}, phase- and velocity-space descriptions.
\newblock {\em Physics Reports}, 185:1--54, 1990.

\bibitem{RovelliObservable}
Carlo Rovelli.
\newblock What is observable in classical and quantum gravity?
\newblock {\em Classical and Quantum Gravity}, 8:297--316, 1991.

\bibitem{RovelliPartialObservables}
Carlo Rovelli.
\newblock Partial observables.
\newblock {\em Physical Review D}, 65:124013, 2002.
\newblock arXiv:gr-qc/0110035.

\bibitem{WeatherallHole}
James~Owen Weatherall.
\newblock Regarding the `hole argument'.
\newblock {\em The British Journal for the Philosophy of Science}, 69:329--350,
  2018.
\newblock arXiv:1412.0303 [physics.hist-ph].

\bibitem{BarbourTimeless1}
Julian~B. Barbour.
\newblock The timelessness of quantum gravity: {I. The} evidence from the
  classical theory.
\newblock {\em Classical and Quantum Gravity}, 11:2853--2873, 1994.

\bibitem{BarbourTimeless2}
Julian~B. Barbour.
\newblock The timelessness of quantum gravity: {II. The} appearance of dynamics
  in static configurations.
\newblock {\em Classical and Quantum Gravity}, 11:2875--2897, 1994.

\bibitem{AndersonProblemofTime}
Edward Anderson.
\newblock {\em The Problem of Time: Quantum Mechanics versus General
  Relativity}.
\newblock Springer, Cham, Switzerland, 2017.

\bibitem{ThebaultCanonicalGRTime}
Karim P.~Y. Th\'{e}bault.
\newblock Three denials of time in the interpretation of canonical gravity.
\newblock {\em Studies in History and Philosophy of Modern Physics},
  43:277--294, 2012.

\bibitem{BelotPOP}
Gordon Belot.
\newblock The representation of time and change in mechanics.
\newblock In Jeremy Butterfield and John Earman, editors, {\em Philosophy of
  Physics, Part A}, Handbook of the Philosophy of Science, pages 133--227.
  Elsevier, Amsterdam, 2007.

\bibitem{GrybThebault}
Sean Gryb and Karim P.~Y. Th\'{e}bault.
\newblock Time remains.
\newblock {\em The British Journal for the Philosophy of Science}, 67:663--705,
  2016.

\bibitem{GrybThebaultSchrodinger}
Sean Gryb and Karim Th\'{e}bault.
\newblock Schrodinger evolution for the universe: Reparametrization.
\newblock {\em Classical and Quantum Graviy}, 33:065004, 2016.

\bibitem{NelsonTeitelboim}
Jeanette~E. Nelson and Claudio Teitelboim.
\newblock Hamiltonian formulation of the theory of interacting gravitational
  and electron fields.
\newblock {\em Annals of Physics}, 116:86--104, 1978.

\bibitem{HenneauxGeometrodynamicsTetrad}
Marc Henneaux.
\newblock On geometrodynamics with tetrad fields.
\newblock {\em General Relativity and Gravitation}, 9:1031--1045, 1978.

\bibitem{HenneauxGeneiauEinsteinDirac}
J.~Geheniau and M.~Henneaux.
\newblock {Einstein-Dirac} equations in suited tetrads.
\newblock {\em General Relativity and Gravitation}, 8:611--615, 1977.

\bibitem{HenneauxBianchiIspinor}
Marc Henneaux.
\newblock Bianchi type-{I} cosmologies and spinor fields.
\newblock {\em Physical Review D}, 21:857--863, 1980.

\bibitem{PonsSalisburyShepleyAshtekar}
J.~M. Pons, D.C. Salisbury, and L.~C. Shepley.
\newblock Gauge group and reality conditions in {Ashtekar's} complex
  formulation of canonical gravity.
\newblock {\em Physical Review D}, 62:064026, 2000.
\newblock arXiv:gr-qc/9912085.

\bibitem{SalisburyBianchiI}
D.~C. Salisbury, J.~Helpert, and A.~Schmitz.
\newblock Reparameterization invariants for anisotropic {Bianchi I} cosmology
  with a massless scalar source.
\newblock {\em General Relativity and Gravitation}, 40:1475--1498, 2008.

\bibitem{AshtekarBianchi}
Abhay Ashtekar and Joseph Samuel.
\newblock Bianchi cosmologies: {The} role of spatial topology.
\newblock {\em Classical and Quantum Gravity}, 8:2191--2215, 1991.

\bibitem{MassiveGravity1}
J.~Brian Pitts and William~C. Schieve.
\newblock Universally coupled massive gravity.
\newblock {\em Theoretical and Mathematical Physics}, 151:700--717, 2007.
\newblock arXiv:gr-qc/0503051v3.

\bibitem{MassiveGravity2}
J.~Brian Pitts.
\newblock Universally coupled massive gravity, {II: Densitized} tetrad and
  cotetrad theories.
\newblock {\em General Relativity and Gravitation}, 44:401--426, 2012.
\newblock arXiv:1110.2077.

\bibitem{MassiveGravity3}
J.~Brian Pitts.
\newblock Universally coupled massive gravity, {III}: {dRGT-Maheshwari} pure
  spin-$2$, {Ogievetsky-Polubarinov} and arbitrary mass terms.
\newblock {\em Annals of Physics}, 365:73--90, 2016.
\newblock arXiv:1505.03492 [gr-qc].

\bibitem{PittsScalar}
J.~Brian Pitts.
\newblock Massive {Nordstr\"{o}m} scalar (density) gravities from universal
  coupling.
\newblock {\em General Relativity and Gravitation}, 43:871--895, 2011.
\newblock arXiv:1010.0227v1 [gr-qc].

\bibitem{Sundermeyer}
Kurt Sundermeyer.
\newblock {\em Constrained Dynamics: With Applications to Yang--Mills Theory,
  General Relativity, Classical Spin, Dual String Model}.
\newblock Springer, Berlin, 1982.
\newblock Lecture Notes in Physics, volume 169.

\bibitem{Wald}
Robert~M. Wald.
\newblock {\em General Relativity}.
\newblock University of Chicago Press, Chicago, 1984.

\bibitem{FradkinVilkoviskyHLEquivalence}
E.~S. Fradkin and G.~A. Vilkovisky.
\newblock Quantization of relativistic systems with constraints: {Equivalence}
  of canonical and covariant formalisms in quantum theory of gravitational
  field.
\newblock Ref.TH 2332.CERN, 1977.
\newblock http://cds.cern.ch/record/406087/.

\bibitem{CastellaniGaugeGenerator}
Leonardo Castellani.
\newblock Symmetries in constrained {Hamiltonian} systems.
\newblock {\em Annals of Physics}, 143:357--371, 1982.

\bibitem{SudarshanMukunda}
E.~C.~G. Sudarshan and N.~Mukunda.
\newblock {\em Classical Dynamics: A Modern Perspective}.
\newblock John Wiley \& Sons, New York, 1974.

\bibitem{BergmannKomar}
Peter~G. Bergmann and Arthur Komar.
\newblock The coordinate group symmetries of general relativity.
\newblock {\em International Journal of Theoretical Physics}, 5:15--28, 1972.

\bibitem{BergmannFadingWorldPoint}
Peter~G. Bergmann.
\newblock The fading world point.
\newblock In Peter~G. Bergmann and Venzo {De Sabbata}, editors, {\em Cosmology
  and Gravitation---Spin, Rotation, Torsion, Rotation, and Supergravity, 1979,
  Erice}, pages 173--176, New York, 1980. Plenum.

\bibitem{BergmannEarly}
Peter~G. Bergmann.
\newblock The canonical formulation of general relativistic theories: The early
  years, 1930-1959.
\newblock In Don Howard and John Stachel, editors, {\em Einstein and the
  History of General Relativity: Based on the Proceedings of the 1986 Osgood
  Hill Conference}, pages 293--299. Birkh\"{a}user, Boston, 1989.

\bibitem{PonsSalisburyShepleyYang}
J.~M. Pons, D.~C. Salisbury, and L.~C. Shepley.
\newblock Gauge transformations in {Einstein-Yang-Mills} theories.
\newblock {\em Journal of Mathematical Physics}, 41:5557--5571, 2000.
\newblock arXiv:gr-qc/9912086.

\bibitem{SuganoKimuraFirstSecond}
Reiji Sugano and Toshiei Kimura.
\newblock Gauge transformations for dynamical systems with first- and
  second-class constraints.
\newblock {\em Physical Review D}, 41:1247--1254, 1990.

\bibitem{ChitaiaGaugeFirstSecond}
N.~P. Chitaia, S.~A. Gogilidze, and Yu.~S. Surovtsev.
\newblock Dynamical systems with first- and second-class constraints. {II.
  Local}-symmetry transformations.
\newblock {\em Physical Review D}, 56:1142--1155, 1997.

\bibitem{ExactSolns2}
Hans Stephani, Dietrich Kramer, Malcolm MacCallum, Cornelius Hoenselaers, and
  Eduard Herlt.
\newblock {\em Exact Solutions of Einstein's Field Equations}.
\newblock Cambridge University Press, Cambridge, second edition, 2003.

\bibitem{OrtinLieLorentzDerivatives}
Tom\'{a}s Ort\'{i}n.
\newblock A note on {Lie-Lorentz} derivatives.
\newblock {\em Classical and Quantum Gravity}, 19(15):L143--L149, 2002.

\bibitem{SzybiakLie}
Andrzej Szybiak.
\newblock On the {Lie} derivative of geometric objects from the point of view
  of functional equations.
\newblock {\em Prace Matematyczne=Schedae Mathematicae}, 11:85--88, 1966.

\bibitem{SzybiakCovariant}
Andrzej Szybiak.
\newblock Covariant derivative of geometric objects of the first class.
\newblock {\em Bulletin de l'Acad\'{e}mie Polonaise des Sciences, S\'{e}rie des
  Sciences Math\'{e}matiques, Astronomiques et Physiques}, 11:687--690, 1963.

\bibitem{Tashiro1}
Yoshihiro Tashiro.
\newblock Sur la d\'{e}riv\'{e}e de {Lie} de l'\^{e}tre g\'{e}om\'{e}trique et
  son groupe d'invariance.
\newblock {\em T\^{o}hoku Mathematical Journal}, 2:166--181, 1950.

\bibitem{Tashiro2}
Yoshihiro Tashiro.
\newblock Note sur la d\'{e}riv\'{e}e de {Lie} d'un \^{e}tre
  g\'{e}om\'{e}trique.
\newblock {\em Mathematical Journal of Okayama University}, 1:125--128, 1952.

\bibitem{BilyalovConservation}
Ranat~F. Bilyalov.
\newblock Conservation laws for spinor fields on a {Riemannian} space-time
  manifold.
\newblock {\em Theoretical and Mathematical Physics}, 90:252--259, 1992.

\bibitem{TAM2013TimeandFermionsConfProc}
J.~Brian Pitts.
\newblock Time and fermions: {General} covariance \emph{vs.} {Ockham's} razor
  for spinors.
\newblock In Martin O'Loughlin, Samo Stani\v{c}, and Darko Veberi\v{c},
  editors, {\em Proceedings of the 4th International Conference on Time and
  Matter, 4-8 March 2013, Venice, Italy}, pages 185--198. University of Nova
  Gorica Press, Nova Gorica, Slovenia, 2013.

\bibitem{BergmannConservation}
Peter~G. Bergmann.
\newblock Conservation laws in general relativity as the generators of
  coordinate transformations.
\newblock {\em Physical Review}, 112:287--289, 1958.

\bibitem{Kiefer3rd}
Claus Kiefer.
\newblock {\em Quantum Gravity}.
\newblock Oxford University Press, Oxford, 3rd edition, 2012.

\end{thebibliography}

\end{document}